\documentclass{egpubl}
\usepackage{eurovis2018}

\SpecialIssuePaper         

 \electronicVersion 

\ifpdf \usepackage[pdftex]{graphicx} \pdfcompresslevel=9
\else \usepackage[dvips]{graphicx} \fi

\PrintedOrElectronic

\usepackage{t1enc,dfadobe}

\usepackage{egweblnk}
\usepackage{cite}

\usepackage{amsfonts}
\usepackage{amsmath}
\usepackage{amssymb}

\title[Cost-benefit Analysis of Visualization in Virtual Environments]%
      {Cost-benefit Analysis of Visualization in Virtual Environments}

\author[M. Chen, K. Gaither, N. W. John, B. McCann]
{\parbox{\textwidth}{\centering %
       Min Chen\thanks{Contact Author: min.chen@oerc.ox.ac.uk}$^{1}$,
       Kelly Gaither$^2$, Nigel W. John$^3$, and Brian McCann$^2$
       }
       \\
{\parbox{\textwidth}{\centering %
         $^1$University of Oxford, UK, \hspace{2mm}
         $^2$University of Texas at Austin, USA, \hspace{2mm} and \hspace{2mm}
         $^3$University of Chester, UK
       } 
 }
 }

\begin{document}

\teaser{
  \vspace{-5mm}
  \centering
  \begin{tabular}{@{}c@{\hspace{2mm}}c@{\hspace{2mm}}c@{\hspace{2mm}}c@{}}
    \includegraphics[height=35mm]{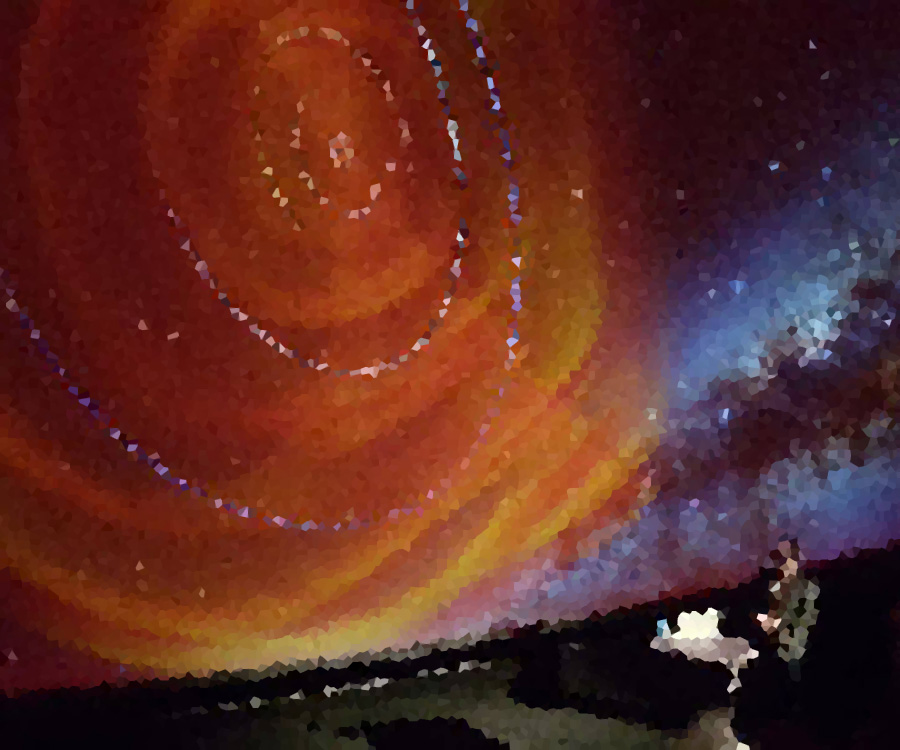} &
    \includegraphics[height=35mm]{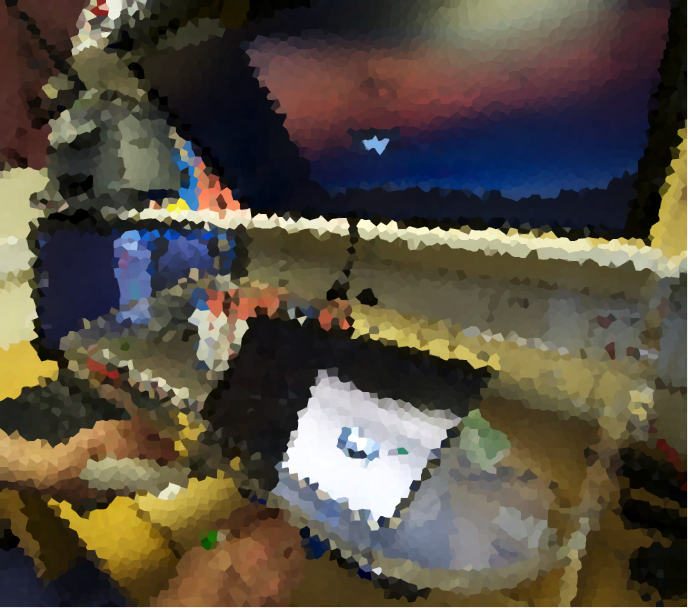} &
    \includegraphics[height=35mm]{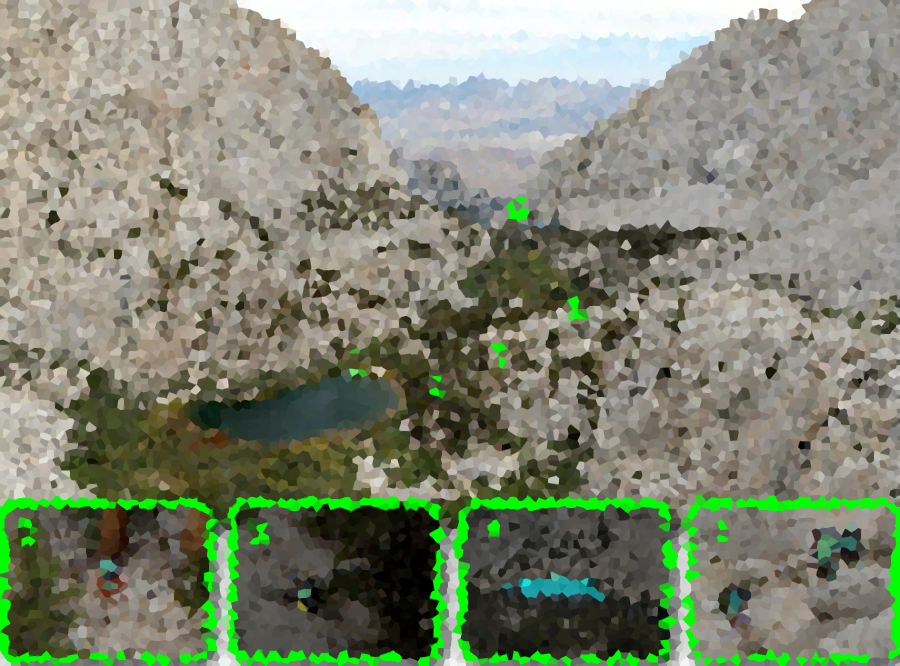} &
    \includegraphics[height=35mm]{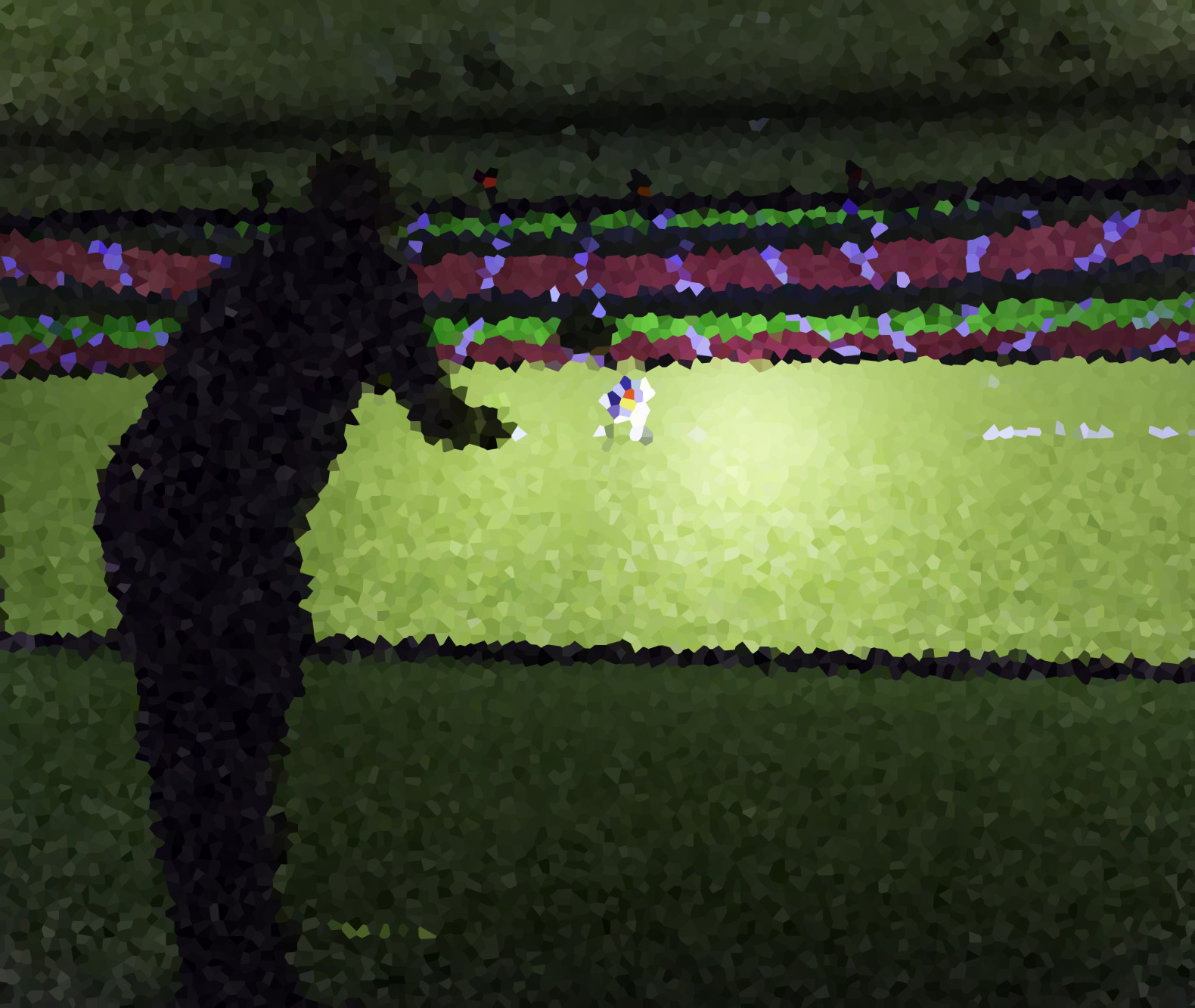} \\
    \small{(a) an OpenSpace event \cite{Bock:2015:SciVisP}} &
    \small{(b) surgical simulation \cite{Villard:2014:CARS}} &
    \small{(c) gigaimage analytics \cite{Ip:2011:TVCG}} &
    \small{(d) sports training \cite{Miles:2014:TCS}}
  \end{tabular}
  \caption{\label{fig:Examples} Four examples of typical virtual environments (VEs) used for visualization applications.
  \emph{Note: The copyrights for these images will be obtained if the paper is accepted for publication.}}
  \vspace{5mm}
}

\maketitle
\begin{abstract}
Visualization and virtual environments (VEs) have been two interconnected parallel strands in visual computing for decades.
Some VEs have been purposely developed for visualization applications, while many visualization applications are exemplary showcases in general-purpose VEs.
Because of the development and operation costs of VEs, the majority of visualization applications in practice are yet to benefit from the capacity of VEs.
In this paper, we examine this perplexity from an information-theoretic perspective.
Our objectives are to conduct cost-benefit analysis on typical VE systems (including augmented and mixed reality, theatre-based systems, and large powerwalls), to explain why some visualization applications benefit more from VEs than others, and to sketch out pathways for the future development of visualization applications in VEs.
We support our theoretical propositions and analysis using theories and discoveries in the literature of cognitive sciences and the practical evidence reported in the literatures of visualization and VEs.


\printccsdesc   
\end{abstract}

\section{Introduction}
\label{sec:Intro}
From a broad perspective, the uses of visualization and virtual environments (VEs) have much in common.
Both facilitate computer-supported activities involving primarily visual perception and human-computer interaction.
Most systems that enable VE research and applications, such as the CAVE (Cave Automatic Virtual Environment) in the 1990s \cite{Disz:1995:book} and the RAVE (Reconfigurable Automatic Virtual Environment) in the 2000s \cite{Brodlie:2005:CGF}, are also considered as large visualization infrastructures.
A variety of visualization applications, ranging from biomedical data visualization to text and document visualization, have been implemented to run in VEs.

Despite the shared common ground, visualization publications rarely feature virtual reality or augmented reality capabilities, while research in VEs seldom addresses commonly understood challenges in information visualization, scientific visualization, or visual analytics.
Concerns regarding the financial return on investment of historical VE hardware, recurring operation and maintenance, and to a lesser degree, software, have in many ways overshadowed the potential values that VEs may have as a viable discovery environment.
Additionally, doubts about the value of conducting visual analytics and sense making in a VE have been a topic of consideration with mixed consensus.
At first glance, the cost-benefit metric for visualization processes proposed by Chen and Golan \cite{Chen:2016:TVCG} indicate that visualization in VEs may suffer from the disadvantages of lacking abstraction and high cost, but a cursory look at the history of VEs and the creativity in this space as a whole suggests there is more to understand. 

In this paper, we investigate the cost-benefit of visualization in VEs from three perspectives: information theory, cognitive sciences, and practical applications.
We use the term \emph{virtual environment} (VE) as an encompassing term for immersive and semi-immersive virtual environments, mixed and augmented reality, visual as well as non-visual perception, and device-based as well as natural interaction.
This investigation serves as a theoretical assessment about the usability of VEs in visualization as well as the applicability of Chen and Golan's cost-benefit metric \cite{Chen:2016:TVCG}.

We frame our discourse based on \emph{immersion} and \emph{presence}, the most fundamental properties of VEs (Section \ref{sec:VEdimensions}).
In the context of \emph{immersion} and \emph{presence}, we first examine the three elementary quantities of the information-theoretic metric for cost-benefit analysis, namely \emph{alphabet compression}, \emph{potential distortion}, and \emph{cost} (Section \ref{sec:Theory}).
We then support the theoretical propositions and analysis with theories and discoveries in the literature of cognitive science (Section \ref{sec:Psychology}).
This is followed by practical evidence reported in the literature of visualization and VEs (Section \ref{sec:Applications}).
Our investigation leads to an in-depth analysis of the cost-benefit of performing four different levels of visualization tasks in VEs.
This analysis enables us to consider the quantitative cost and benefit of \emph{immersion} and \emph{presence} at each level.
It offers theory- and evidence-based explanations of the past implementations, while suggesting new opportunities and challenges.

Our contributions include (i) the application of theory- and evidence-based cost-benefit analysis to an important but often overlooked area of visualization (Sections \ref{sec:Theory}--\ref{sec:Applications}), (ii) a collection of fundamental discoveries about the merits and demerits of performing visualization tasks in VEs (Section \ref{sec:Vision}), and (iii) a demonstration that the theory can guide us to explore answers to practical questions (Section \ref{sec:QA}).
In particular:

\begin{itemize}
\item
The increase of \emph{presence} leads to the increase of \emph{attention} and in some cases \emph{enjoyment}, which is desirable to the presenter in Disseminative Visualization (Level 1).
\item
The increase of \emph{presence} leads to the reduction of \emph{potential distortion} by making use of humans' \emph{memory} and \emph{a priori knowledge}, which is desirable in some Observational Visualization (Level 2) where perceived information must be \emph{associated} with reality efficiently and effectively.
\item
The increase of \emph{presence} leads to the increase of \emph{alphabet compression} and the reduction of \emph{potential distortion} and \emph{learning cost} in some Model-Developmental Visualization (Level 4) where human participants' \emph{behavioral models} can be studied, and humans' learning capabilities can be utilized.
\item
The increase of \emph{presence} usually leads to the decrease of \emph{alphabet compression} and increase of \emph{cost}, which is often undesirable in Observational and Analytical Visualization (Level 2 and Level 3), especially when non-intuitive mapping (not easy to learn and remember) from data to virtual objects is deployed.
\item
The increase of \emph{presence} usually leads to the increasing demand for \emph{attention}, and failing to meet such \emph{cost} often leads to \emph{inattentional blindness} in visualization.
\item
Analytical visualization tasks and (algorithmic) model-developmental tasks typically present a large and complex search space for the target patterns or optimized solutions. The increase of \emph{immersion} and \emph{presence} has potential to provide a means to explore a large and complex search space. We also touch briefly on an open question: How can we introduce intuitive and effective \emph{presence} to support humans' \emph{intelligence} in discovering target patterns or optimized solutions?
\end{itemize}

\section{Related Work}
\label{sec:RelatedWork}
In this paper, we use the term \emph{Virtual Environments} (VEs) as an encompassing term for \emph{immersive} and \emph{semi-immersive} environments, large theatre- or dome-based infrastructures, gigaimage displays, virtual reality systems, mixed reality systems, augmented reality systems, augmented virtuality systems, and web-based VEs.
There are numerous VE systems and applications reported in the literature.
Readers who are interested in exploring the broad spectrum of VEs may consult a number of books and literature surveys on the subject
\cite{Slater:2001:book,Vince:2013:book}
as well as in specific areas, including, but not limited to,
presence \cite{Schuemie:2001:CPB,Slater:2013:report},
haptics \cite{Coles:2011:H},
augmented reality \cite{Azuma:2001:CGA,Sielhorst;2008:JDT},
usability evaluation \cite{Bowman:2002:P},
medicine and healthcare \cite{Alaraj:2011:SNI,Cosentino:2013:JRP,Weiss:2016:book},
flight simulation \cite{Lee:2005:book,Rolfe:2008:book},
education \cite{Bell:2016:book},
sports \cite{Miles:2012:CG},
and cultural and natural heritage \cite{Addison:2000:M}.

Milgram et al. outlined the \emph{Reality-Virtuality Continuum} ~\cite{Milgram:1995:PIA} that defines a continuous scale ranging between the completely virtual and the completely real.
The area between these two extremes is referred to as \emph{mixed reality}, which encompasses the technology of \emph{augmented reality} where the virtual augments the real and the technology of \emph{augmented virtuality} where the real augments the virtual.
Schnabel et al. enriched this continuum by relating the correction between action and perception to the extent of interaction with real objects \cite{Schnabel:2007:IASDR}.
In this work, we will explore this continuum by examining the cost-benefit of virtuality and reality in visualization processes.

The theoretical research in the field of VEs has been largely focused on the concept of \emph{presence}.
Researchers have engaged in extensive discourse as to what is the sense of presence and what may contribute to such a sense.
Sheridan \cite{Sheridan:1992:P} and Heeter \cite{Heeter:1992:P} are among the first to initiate this discourse.
Sheridan \cite{Sheridan:1992:P} outlined three causes of presence: the extent of sensory information, the control of the relation between sensors and an environment, and the ability to modify a physical environment.
Heeter \cite{Heeter:1992:P} drew distinction between three types of presence, namely personal, social, and environmental presence.
Schloerb \cite{Schloerb:1995:P} divides the notion into two categories, subjective and objective presence.
Slater and Wilbur \cite{Slater:1997:P} related these two categories to two distinctive terms \emph{presence} and \emph{immersion} respectively.
Lombard and Ditton \cite{Lombard:2000:P} defined six aspects of presence: social richness, realism, transportion, immersion, social actor with medium, and medium as social actor.
Stater et al. \cite{Slater:2013:report} further defined the dimensions of \emph{presence} and \emph{immersion}.
Schuemie et al. gave a comprehensive review about this line of inquiry \cite{Schuemie:2001:CPB}.
In this paper, we relate the concepts of presence and immersion to the abstract properties of \emph{alphabet compression}, \emph{potential distortion}, and \emph{cost} in visualization processes \cite{Chen:2016:TVCG}.
We examine when and where presence and immersion may be beneficial to visualization users, and when and where they incur a noticeable amount of costs.

The theoretical research in the field of visualization has resulted in
a large number of taxonomies (e.g., \cite{Bertin:1983:book,Tory:2004:InfoVis}),
many conceptual models (e.g., \cite{vanWijk:2005:Vis,Moreland:2013:TVCG}), and
a few theoretic frameworks (e.g., \cite{Chen:2010:TVCG,Xu:2010:TVCG,Kindlmann:2014:TVCG}).
A more comprehensive list of references can be found at \cite{Chen:2017:web}.
Recently Chen et al. \cite{Chen:2017:CGA} suggested that the theoretical foundation of visualization includes four major aspects, namely \emph{taxonomies and ontologies}, \emph{principles and guidelines}, \emph{conceptual models and theoretic frameworks}, and \emph{quantitative laws and theoretic systems}.
This work falls into the category of conceptual models and theoretic frameworks.
We aim to use information theory \cite{Shannon:1948:BSTJ} to bring a substantial amount of activities in VEs into the information-theoretic framework for visualization \cite{Chen:2016:book}.
Once visualization activities in VEs can be considered as data intelligence processes, we can categorize these activities based on the four levels of visualization tasks \cite{Chen:2016:TVCG}, and apply information-theoretic metric for cost-benefit analysis to these activities in an abstract and objective manner.
Meanwhile, this work also provides an opportunity to evaluate the theoretic findings in \cite{Chen:2016:TVCG} to see if it can explain complex phenomena in visualization and VEs, if its analytical discourse can be supported by evidence in cognitive sciences and real-world applications, and if it can be used to suggest new guidelines, hypotheses, and predictions.

There has always been an interest in VEs in the field of visualization.
For example, in 1995, Disz et al. \cite{Disz:1995:book} reported visualization experience in a CAVE, and Taylor et al. \cite{Taylor:1995:book} presented performance models for interactive and immersive visualization for scientific applications.
In recent years, Ip et al. \cite{Ip:2011:TVCG} reported the use of a large VE system for gigapixel analytics.
Reda et al. \cite{Reda:2013:CGA} reported the use of CAVE2 for visualizing large, heterogeneous data.
Bock et al. \cite{Bock:2015:SciVisP} showcased a dome-based VE infrastructure for Open Science events.
Papadopoulos et al. \cite{Papadopoulos:2015:CGA} presented an immersive gigapixel display, and Papadopoulos and Kaufman \cite{Papadopoulos:2013:TVCG} presented techniques that enable focus-and-context visualization using such a system.
M\"{u}ller et al. \cite{Muller:2016:IJSI} presented an evaluation of biological data visualization using a large VE system.
We hope that this work will stimulate new interests in delivering visualization solutions using VEs.

\begin{figure*}[t!]
  \centering
  \includegraphics[width=\linewidth,height=35mm]{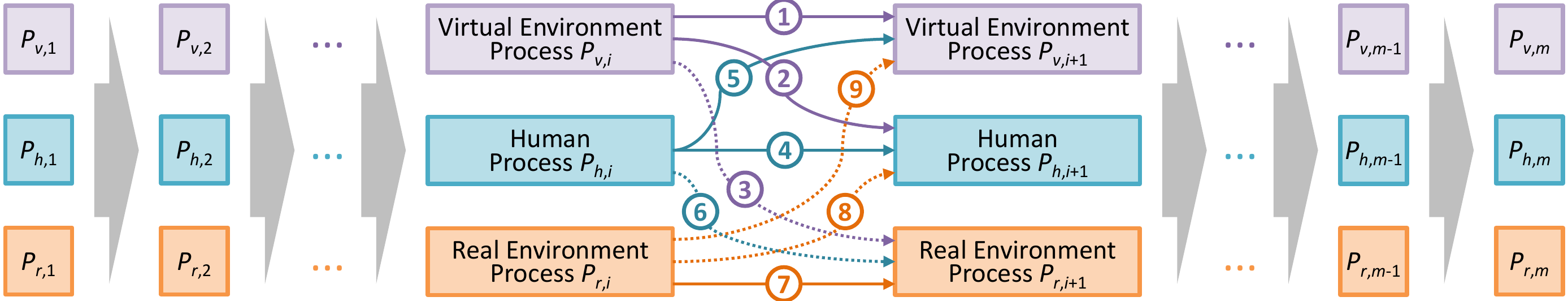}
  \caption{\label{fig:ProcessFlow}%
A sequence of events in a VE can be considered as a series of processes flowing along a pathway in a complex space of all possible states of the entities involved. The main entities are the system of the VE and the human participant(s). In a mixed reality environment, parts of the reality are also involved as the third entity. While such processes result in changes at each stage, information is passed from the processes at stage $i$ to those at stage $i+1$ along paths marked by \textcircled{\small{1}}-\textcircled{\small{9}}. The dotted lines indicate those paths typically available only in mixed reality environments.
}
\vspace{-6mm}
\end{figure*}

\section{Dimensions of Virtual Environments}
\label{sec:VEdimensions}
Research in virtual environments (VEs) differs from that in computer graphics and visualization by placing a significant emphasis on the concepts of \emph{immersion} and \emph{presence}.
While many have contributed to the formulation of these two concepts, we chose to adopt Slater et al.'s definitions \cite{Slater:2013:report} as a basis for our investigation.

\noindent\textsf{\textbf{Immersion}} is an attribute used to describe a technology.
It characterizes the extent to which a VE is capable of delivering an inclusive, extensive, surrounding, and vivid illusion of reality to the senses of a human participant.
There are six dimensions of immersions described in \cite{Slater:2013:report}:
\begin{itemize}
\item \emph{inclusion} -- the extent to which physical reality is shut out,
\item \emph{extension} -- the range of sensory modalities accommodated,
\item \emph{surrounding} -- the extent of visual coverage (e.g., panoramic, telescopic, microscopic, x-vision, etc.);
\item \emph{vividness} -- the fidelity of the information conveyed (e.g., display resolution, color resolution, content richness, and variety of energy simulated within a particular modality);
\item \emph{match} -- the degree of correlation between the information conveyed by the VE and a participant's proprioceptive feedback about body movements; and
\item \emph{plot-interactivity} -- the extent to which a participant can influence the storyline or the sequence of events in a VE.
\end{itemize}
These dimensions of immersion are considered to be measurable objectively and quantitatively.
There have been a number of experiments designed to obtain these measurements for specific VEs.

\noindent\textsf{\textbf{Presence}} is an attribute used to describe a human participant.
It characterizes the state of consciousness, i.e., the psychological sense of being in a VE.
In contrast to immersion, describing presence is often subjective and qualitative in nature, although some aspects of presence may be measurable objectively and quantitatively.
The state of consciousness can be described by, but not limited to, the following senses:
\begin{itemize}
\item a sense of believing \cite{Sheridan:1992:P}, e.g., being at a place vs. viewing a set of images,
\item a sense of naturalism \cite{Bowman:2012:CACM}, e.g., acting as if in the real-world vs. acting unnaturally,
\item a sense of social presence \cite{Short:1976:book}, e.g., participating in face-to-face interaction vs. remote communication,
\item a sense of co-presence \cite{Nowak:2001:P}, e.g., being together with other actors vs. unconnected individual actors.
\end{itemize}
In general, the most technical advances in VEs have been driven by higher specifications of immersion and increased requirements for presence.
Comparing a VE featuring more immersion or presence with a VE featuring less, the former generally delivers more data to a user through its its available information channels (visual, audio, etc.).
However, the former often incurs more costs, leading naturally to a question regarding the cost-benefit of different VEs from an information-theoretic perspective.

\section{Theoretic Propositions and Analysis}
\label{sec:Theory}
\noindent \textsf{\textbf{Processes and States.}}
A sequence of interactive events in a VE can be considered as a processing flow as illustrated in Figure \ref{fig:ProcessFlow}.
In most VEs, there are two main types of processes: \emph{virtual environment (VE) processes} and \emph{human} processes.
\emph{VE Processes} include all machine-centric processes that enable the devices in a VE to change their states, e.g., generating new images, sounds, or force-feedback functions.
\emph{Human Processes} are human-centric and encompass any processes that enable the participants in a VE to change their states, e.g., attention, perception, interpretation, memory, emotion, speech, and body actions.
In mixed reality environments, including augmented reality and augmented virtuality, a participant's reality may also change.
We refer to the causes of such changes as \emph{Real Environment Processes}.

In theory, the steps in Figure \ref{fig:ProcessFlow} can be infinitesimally small in time, the resulting changes can be infinitesimally detailed, while the sequence can be innumerably long and the processes can be immeasurably complex.
In practice, one can construct a coarse approximation of a processing flow for a specific set of tasks.
We will adopt this approach when we examine practical VE systems.

We can finely divide the time steps, and the interaction among the three classes of processes can be defined as forward connections as shown in Figure \ref{fig:ProcessFlow}.
The connections \textcircled{\small{1}}, \textcircled{\small{4}}, and \textcircled{\small{7}} indicate the state transitions within the same class of processes.
A VE system that delivers output at time $t_{i+1}$ is expected to know the state at time $t_i$.
The position of a human participant at time $t_{i+1}$ is expected to be caused by a movement from the corresponding position at time $t_i$.

Meanwhile, a human participant can receive a variety of information conveyed by the VE system as indicated by connection \textcircled{\small{2}}, and machine-sensors can pick up aspects of a human state as indicated by connection \textcircled{\small{5}}.
In a mixed reality environment, information is also passed from the real environment processes to the human participant and machine sensors as indicated by \textcircled{\small{8}} and \textcircled{\small{9}}.
When an object in the real environment is manipulated by a human participant or a device in the VE (e.g., a robotic arm), we conceptualize this phenomenon as information communication from a human process or a VE process to a real environment process.

We note that all of these processes receive information from a previous state, process the information, and deliver changes as a transformation to a new state.
This processing flow bears a remarkable resemblance to a data analysis and visualization workflow \cite{Chen:2016:TVCG}.
Furthermore, the concept of \emph{immersion} is a collective and accumulative attribute primarily about the machine-centric VE processes, while the concept of \emph{presence} is a collective and accumulative attribute primarily about the human-centric processes.
Hence, it is not surprising that VEs have been used for visualization applications.
While VEs have emerged in the consumer gaming market, for the purposes of this paper, we consider only visualization applications and their function in virtual environments.

\begin{table*}[t!]
  \centering
  \caption{\label{tab:VEsystems}%
The design emphases of some typical VE systems, and their abstraction in terms of the cost-benefit measures.}
  \includegraphics[width=\linewidth]{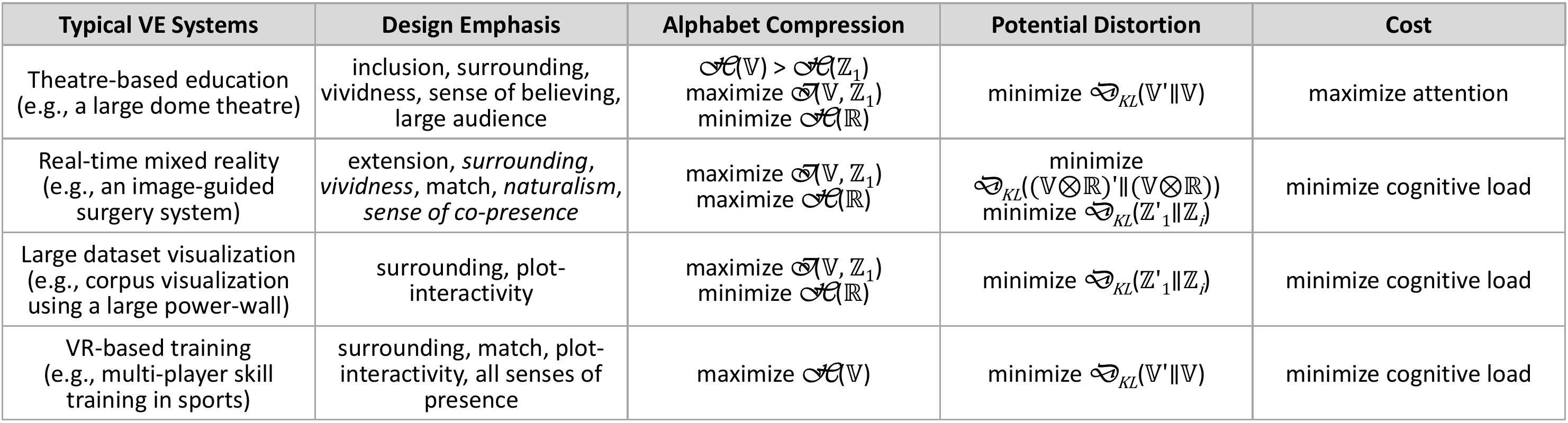}
\vspace{-8mm}
\end{table*}

\noindent \textsf{\textbf{Alphabets and Letters.}}
In abstract, let all possible states of VE processes (e.g., combinations of different computer-generated scenes, sounds, force-feedbacks, etc.) be the \emph{letters} of a very large \emph{alphabet} $\mathbb{V}$,
all possible states of human processes (e.g., combinations of the physical and cognitive states of all human participants in a VE) be the \emph{letters} of a very large \emph{alphabet} $\mathbb{H}$, and
all possible states of the reality observable to the VE system and the human participants in the VE be the \emph{letters} of a very large \emph{alphabet} $\mathbb{R}$.%
\footnote{It is helpful to note that the abundance of these letters and the complexity of these alphabets should not be the reason to shelve a theatrical notion.
In the history of thermodynamics, which information theory is rooted, the kinetic theory, which models a gas based on the probabilistic behaviors of a huge number of particles, was difficult to appreciate before 1900s.}
Therefore, the change from one state to another is the same as the change from one letter to another.

Because the variables for these states remain more or less the same in a processing flow, we can maintain the same set of letters in each alphabet (i.e., $\mathbb{V}$, $\mathbb{H}$, or $\mathbb{R}$) in the processing flow, but allow the probabilities of its letters to vary from one moment to another.
For example, a participant may have a state of ``fallen on the floor''.
Although this state may not happen in every session, it can still be included as a letter in the alphabet $\mathbb{H}$.
Its probability varies depending on the task a participant is performing, the mobility skill of a participant, and other factors.
One observation that we can make is that the Shannon entropy of each of the three alphabets in VEs does not have a general trend of reduction along the processing flow.
Moreover, any increase of immersion and presence will most likely result in an increase of the size and complexity of alphabet $\mathbb{V}$, hence an increase of the Shannon entropy of $\mathbb{V}$.
This is not typical in a conventional data analysis and visualization workflow as observed in \cite{Chen:2016:TVCG}.

However, when considering a visualization application in a VE, there is another series of transformation of alphabets, i.e., from a data alphabet at the beginning to a decision alphabet at the end.
Here we refer to these alphabets, which are denoted as $\mathbb{Z}_1, \mathbb{Z}_2, \ldots, \mathbb{Z}_n$, collectively as \emph{visualization alphabets}.
Unlike alphabet $\mathbb{V}$, these visualization alphabets may differ significantly in terms of data type or data resolution.
Some of these alphabets, such as visualization images, will be a constituent part of the VE alphabet $\mathbb{V}$.
But others, such as humans perception about various data patterns, will be part of the human alphabet $\mathbb{H}$.
In a mixed reality environment, some of these alphabets will be a constituent part of the real environment alphabet $\mathbb{R}$.
It is not difficult to imagine that in some cases, the availability of the reality, i.e., letters in $\mathbb{R}$ are limited or too costly; hence one uses aspects of a VE alphabet $\mathbb{V}$ to simulate these letters in $\mathbb{R}$.
In other cases, the desired immersion and presence cannot be achieved entirely using a VE alphabet $\mathbb{V}$ or it is too costly to achieve; hence one mixes some aspects of a reality alphabet $\mathbb{R}$ with those of $\mathbb{V}$.
A fundamental question is: since VEs normally cost more than an everyday visualization environment, what is the benefit that would justify the extra cost?

\noindent \textsf{\textbf{Cost-benefit Analysis.}}
If the observation in \cite{Chen:2016:TVCG} can be applied to VEs, the visualization alphabets, $\mathbb{Z}_1, \mathbb{Z}_2, \ldots, \mathbb{Z}_n$, should also exhibit a general trend of \emph{Alphabet Compression}, since the decision alphabet is usually much smaller than the original data alphabet in terms of Shannon entropy.
In addition to alphabet compression, the cost-benefit metric for optimizing visualization processes \cite{Chen:2016:TVCG} also includes two other abstract measures, \emph{Potential distortion} and \emph{Cost} as shown in Eq.\,(\ref{eq:CBR}):
\begin{equation}
\label{eq:CBR}
   \frac{\text{Benefit}}{\text{Cost}} = \frac{\text{Alphabet Compression}-\text{Potential Distortion}}{\text{Cost}}
\end{equation}
\noindent The equivalent mathematical formulation of Eq.\,(\ref{eq:CBR}) can be found in \cite{Chen:2016:TVCG}.
This metric suggests several principles in data intelligence. For example, \emph{Alphabet Compression} has a positive impact as long as \emph{Potential Distortion} or \emph{Cost} is not increasing.
Human knowledge can reduce the \emph{Potential Distortion} and \emph{Cost} in reconstructing data from visualization.
The \emph{Cost} reflects economic, cognitive, and other types of resources.
Below we examine the dimensions of several typical VE systems, and relate these dimensions to the three abstract components in the metric.
Table \ref{tab:VEsystems} summarizes the above four types of VEs in terms of their design emphases (i.e., dimensions of immersion and presence), and the corresponding measures in the cost-benefit metric.
We elaborate on each below.

\subsection{Theatre-based Education Systems}
\label{sec:Education}
Many visualization applications in VEs are designed for educational purposes, and they are a form of \emph{disseminative visualization} \cite{Chen:2016:TVCG}.
They typically run in conjunction with a theatre-based setup, which can accommodate tens to hundreds of participants.
The large number of participants pose challenges in some dimensions of immersion and presence, such as \emph{extension}, \emph{plot-interactivity}, \emph{social presence}, and \emph{co-presence} as described in Section \ref{sec:VEdimensions}.
Their design mostly focuses on the following dimensions:
\begin{itemize}
\item \emph{inclusion}, e.g., using a very dark theatre to block out reality;
\item \emph{surrounding}, e.g., using a large panoramic display featuring many more pixels than a typical commodity display screen;
\item \emph{vividness}, e.g., using high quality computer-generated imagery resulting from high resolution modelling and sophisticated rendering techniques such as global illumination; and
\item \emph{sense of believing}, e.g., seeing a black hole as a phenomenon as if it is observable to naked eyes.
\end{itemize}
Consider that the VE alphabet $\mathbb{V}$ includes primarily the data being visualized, visual imageries, commentary voice, and accompanying music.
The initial visualization alphabet (i.e., the data alphabet) $\mathbb{Z}_1$ is a subset of $\mathbb{V}$.
Hence, ideally, the mutual information $\mathcal{I}(\mathbb{V}; \mathbb{Z}_1)$ between the two alphabets should be maximized, and roughly the same amount of the entropy of $\mathbb{Z}_1$.
The additional visual and audio effects result in additional entropy $\mathcal{H}(\mathbb{V}) - \mathcal{H}(\mathbb{V}|\mathbb{Z}_1)$.
Meanwhile, the design emphasis on \emph{inclusion} implies the minimization of the entropy of the reality $\mathcal{H}(\mathbb{R})$.

The final visualization alphabet (i.e., the decision alphabet) $\mathbb{Z}_n$ is vaguely defined in such VEs.
The participants are expected to absorb as much information as possible.
This can be defined as the minimization of the potential distortion when a participant remembers the VE alphabet $\mathbb{V}$ as a reconstructed alphabet $\mathbb{V}'$.
The potential distortion is defined as the Kullback-Leibler divergence $\mathcal{D}_{KL}(\mathbb{V}'||\mathbb{V})$.
When the decision alphabet $\mathbb{Z}_n$ is not precisely defined, we can also define the minimization of the potential distortion as the maximization of the mutual information between the final takeaway messages and the VE alphabet, $\mathcal{I}(\mathbb{Z}_n; \mathbb{V})$.
If the decision alphabet $\mathbb{Z}_n$ is relatively small and clearly defined, e.g., understanding the results of an election, the mutual information will be small.
The advantage of maintaining a large and complex alphabet $\mathbb{V}$ throughout a processing flow will disappear.

Interestingly, this type of VE purposely demands a huge amount of cognitive attention from the participants throughout a processing flow.
This demand is facilitated by several immersion dimensions such as \emph{inclusion}, \emph{surrounding}, and \emph{vividness}.
The participants in the VE bear the responsibility to absorb as much information as possible, generally assumed to be an acceptable responsibility.
Because such cognitive load is the cost borne by the participants, the provider of the disseminative visualization does not bear this cost.

However, as mentioned previously, there are large facility and operation costs paid by the VE provider.
In some cases, such as the London Planetarium, the financial costs are partly or fully covered by the participants as an entrance fee.
In other cases, governments and private sponsors are able to fund these types of educational activities as a good cause.
The VEs that have an entry charge implicitly assume a significantly higher cost for providing the participants with a novel experience of accessing information.
Meanwhile, for the information provider, the quality of immersion and presence is of utmost importance to eliminate potential distractions that would divert participants' cognitive load to other tasks.

\subsection{Real-time Mixed Reality Systems}
\label{sec:MixedReality}
Many mixed reality systems are designed to support the needs for real-time visualization.
For example, given an initial dataset (e.g., a computed tomography scan, or a planned route on a map), a mixed reality system may enable the user to visualize the data in conjunction with aspects of the reality (e.g., a patient, or a landscape in the real world).
The visualization tasks are usually reasonably well-defined, e.g., verifying the position of an anatomical feature shown in the visualization against the actual geometry of the patient, or determining the geographical features in the landscape that correspond to the planned route.
These are typical \emph{observational visualization} tasks.
Performing such a task falls neatly into the visualization workflows discussed in \cite{Chen:2016:TVCG}.
The initial dataset is a letter in the data alphabet (i.e., $\mathbb{Z}_1$), and the visualization tasks are represented by the decision alphabet (i.e., $\mathbb{Z}_n$).

In a perfectly idealized situation, one might wish to have the relevant aspects of the reality (e.g., the patient or the landscape) captured as a high-resolution 3D model by a computer system, and the captured reality could then be visualized using high-fidelity rendering in conjunction with the dataset.
In other words, the VE alphabet $\mathbb{V}$ will include aspects of the reality as well as the data.
However, the current technological limitation gives rise to many problems.
For example, the relevant aspects of the reality might change dynamically, and any captured 3D model would become unsynchronized with the reality almost immediately after its capture.
The computational costs for processing a high-resolution 3D model and rendering high-fidelity visualization could be incompatible with the real-time task requirement and the operational environment.
A low-resolution model or low-fidelity visualization of the model would incur more cognitive load of the user in relating the visualization to the reality.

A mixed reality system addresses the aforementioned technological limitation by introducing the reality as part of the visualization solution.
In terms of immersion and presence, the real environment alphabet $\mathbb{R}$ delivers a substantial amount of the requirements for \emph{extension}, \emph{surrounding}, \emph{vividness}, \emph{plot-interactivity}, and the senses of \emph{believing}, \emph{naturalism}, \emph{social presence}, and \emph{co-presence}.
The main technical challenge is with aspects of \emph{match} between the true reality and the perceived information through viewing the integrated visual representation of data $\mathbb{V}$ (assuming $\mathbb{Z}_1 \subseteq \mathbb{V}$) and $\mathbb{R}$.

In terms of cost-benefit analysis, the alphabet compression from $\mathbb{V} \cup \mathbb{R}$ to the decision alphabet $\mathbb{Z}_n$ is expected to be very high.
The potential distortion depends on the immersion attribute \emph{match}, which can be influenced by many factors, such as the capability of the mixed reality system and the user's experience in registering $\mathbb{V}$ against $\mathbb{R}$.
In comparison with the idealized VE system that captures all required aspects of the reality $\mathbb{R}$, the mixed reality approach is likely to be more economic in the short to medium term.
The potential distortion due to deficiencies in achieving adequate \emph{match} can be alleviated by having more information in the VE alphabet $\mathbb{V}$ about the reality.
The more overlapping between the virtuality $\mathbb{V}$ and the reality $\mathbb{R}$, the more mutual information $\mathcal{I}(\mathbb{V}; \mathbb{R})$,
and the less potential distortion.
Hence, if the technologies for capturing some aspects of reality are becoming more usable and less costly, it is possible to increase the amount of data for representing the reality, such as a 3D model of a patient captured using camera or a 3D landscape captured using drones.
If such 3D models were captured prior to the real-time visualization, they could potentially be used to enhance the user's ability to \emph{match} the virtuality with the reality, while reducing the cognitive load in registering $\mathbb{V}$ against $\mathbb{R}$.

\subsection{``Big Data'' Visualization Systems}
\label{sec:BigData}
Many large VE infrastructures are equipped with gigapixel displays.
Typically, they have been used for visualizing some very large datasets \cite{Reda:2013:CGA}, such as gigapixel images \cite{Ip:2011:TVCG}, and large biomolecular models for simulating the dynamic behaviors of millions of atoms \cite{Muller:2016:IJSI}.
The visualization tasks involved usually fall into any of the four levels of visualization \cite{Chen:2016:TVCG}.
For example, creating an archeological exhibition \cite{Barcelo:2000:book} is a \emph{disseminative visualization} task.
Interactive exploring multi-gigapixel images for object identification \cite{Ip:2011:TVCG} is an \emph{observational visualization} task.
Identifying patterns in social networks is an \emph{analytical visualization} task.
Validating or debugging a large biomolecular model is a \emph{model-developmental visualization} task.

Because the applications in question often do not demand presence dimensions, such as the sense of \emph{believing}, \emph{naturalism}, or \emph{social presence}, these VEs usually place less emphasis on some immersion dimensions such as \emph{inclusion}, \emph{extension}, and \emph{match}.
They are sometimes referred to as \emph{semi-immersive environments}.
The requirements for displaying ``big data'' naturally leads to an emphasis on \emph{surrounding} and \emph{vividness}.
In most cases, the users are given a substantial amount of control, hence a high-level of \emph{plot interactivity}.
In many applications, the VEs allow multiple users to perform their visualization tasks collaboratively, though the sense of \emph{co-presence} usually arrives naturally through the reality rather than through the display media and interaction devices of a VE.

The relative merits and demerits of using a gigapixel display in comparison with using one or a few conventional desktop displays (referred to as megapixel displays) are always a concern in the minds of many technology providers and users.
We can consider the potential merits and demerits using the information-theoretic metric for cost-benefit analysis.

For observational, analytical, and model-developmental visualization tasks, the VE alphabet $\mathbb{V}$ usually focuses on the data alphabet $\mathbb{Z}_1$.
The visualization tasks are expected to be the same for a gigapixel display and a few megapixel displays.
The decision alphabet $\mathbb{Z}_n$ is thus the same.
Both types of displays are expected to deliver the same amount of alphabet compression.

For a ``big data'' application, the entropy of $\mathbb{Z}_1$ will be very high.
The entropy of a dataset is not necessarily defined by the size of the dataset.
It represents the uncertainty about the amount of potential variations in an alphabet.
In a specific application context (e.g., landscape images in \cite{Ip:2011:TVCG}), the larger the dataset, the more potential variations, and therefore the greater the entropy.
Every time, a user observes a portion of the dataset, some uncertainty disappears.
Comparing a gigapixel display against a megapixel display, the costs for a user to observe the visualization of a very large dataset include the number of interactions, the amount body movement, and the imposition for using the equipment (e.g., finance, inconvenience, etc.).
In general, we expect a megapixel display requires more interaction, a gigapixel display requires more body movement, and encounters more imposition.

One major cognitive difficulty in observing a very large visualization is the need to remember what was observed a moment ago.
With a gigapixel display, this means revisiting a portion of the display with a quick glance at a distance or by walking back to have another close look at the portion again.
With a megapixel display, this typically means relocating the portion concerned through a sequence of interactions, which may not always be straightforward.
This may inevitably result in poor external memorization, cause some potential distortion, and incur additional cognitive load.

Hence, the type of applications that can benefit from a gigapixel display features datasets unfamiliar to a user, with uncertainty (i.e., potential variations) across different parts of a very large visualization, at different resolutions (i.e., zoom factors).
The user's prior knowledge about the dataset and its visualization usually reduces the cost-benefit of using a gigapixel display.

\begin{figure}[t!]
  \centering
  \includegraphics[width=\linewidth]{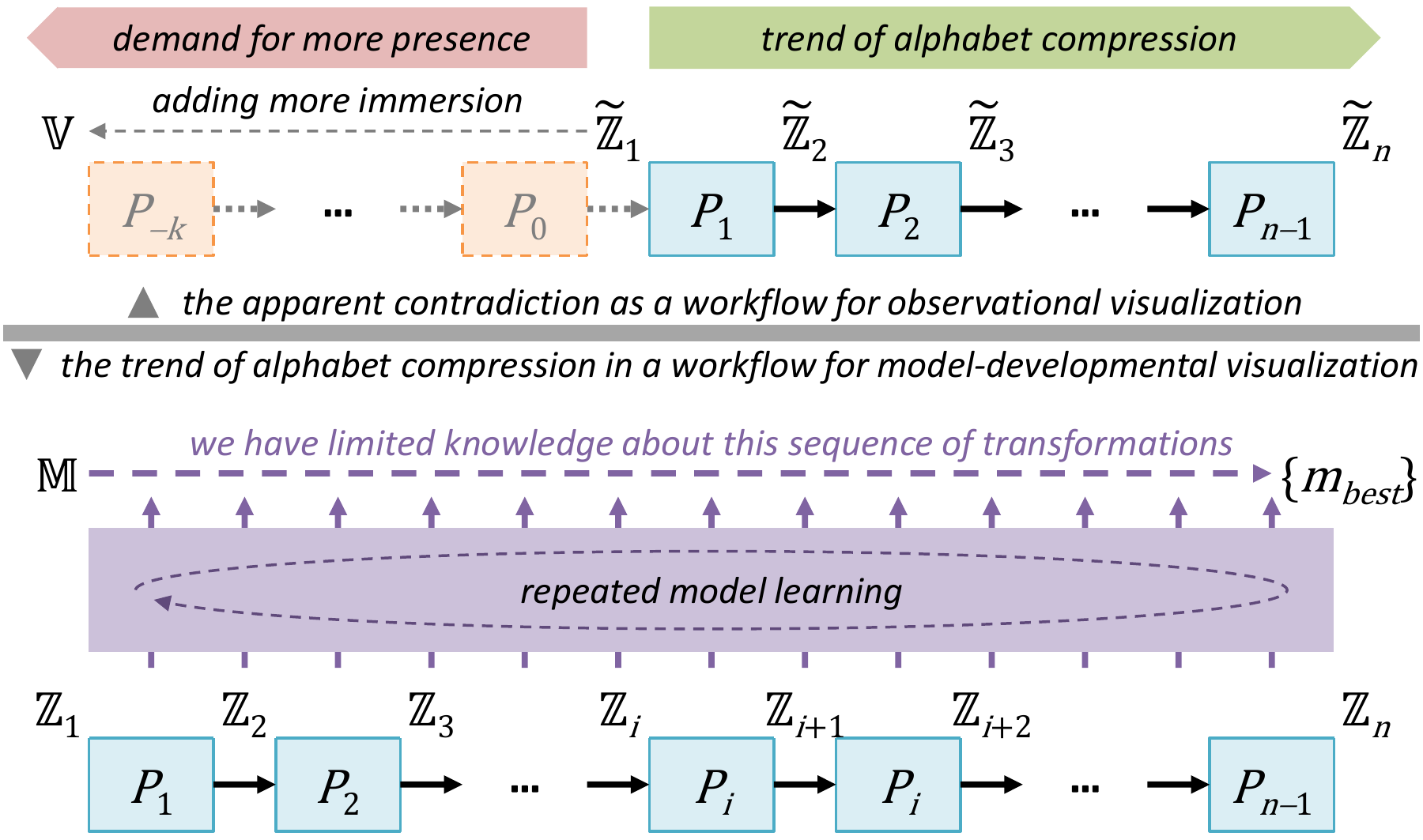}
  \caption{\label{fig:Training}%
If a VE-based training system were considered as a workflow for observational visualization (above), there would be a contradiction with the cost-benefit metric \cite{Chen:2016:TVCG} for optimizing visualization processes. However, it is more appropriate to consider a VE-based training system as a workflow for model-development visualization (below). Because we have limited knowledge about the structure of the model, the variables that affect the model, and the evolution of the model, the VE tends to maximize the amount of reality that can be simulated.
}
\vspace{-6mm}
\end{figure}

\subsection{VE-based Training Systems}
\label{sec:Training}
One major application of VEs is training, for example, in medicine and sports.
The basic workflow for VE-based training involves repeated exercises where a user receives various stimuli and responds the stimuli with appropriate actions.
In most cases, the stimuli are visual imagery, and the actions are the user's motions or interactions.
While different applications may place emphasis on different dimensions of immersion and presence, all these dimensions normally have a positive role to play if they can be made available.

The primary reason for using VE-based training is the lack of access to the required reality $\mathbb{R}$.
For example, it would be inappropriate to train certain medical procedures on real patients; it would be foolish to set fires on many arbitrary buildings in order to train firefighters; and it would be costly to create many different scenarios in sports using real players.
Hence, one creates a VE alphabet $\mathbb{V}$ to approximate $\mathbb{R}$.
The desired variations in $\mathbb{R}$ are stimulated by different letters in $\mathbb{V}$.

Let us focus on visual stimuli, and consider the VE-alphabet $\mathbb{V}$ as the data alphabet $\widetilde{\mathbb{Z}}_1$ and all the possible actions in response to the visual stimuli as a decision alphabet $\widetilde{\mathbb{Z}}_n$.
Here we use $\tilde{\mathbb{Z}}$ to indicate that this is a very rough approximation, and the actual $\mathbb{Z}$ is more complex as discussed below.
There is a trend in alphabet compression from $\widetilde{\mathbb{Z}}_1$ to $\widetilde{\mathbb{Z}}_n$, and in many training applications, the transformations may take a split second.

This seems to suggest an inconsistency with the theory proposed in \cite{Chen:2016:TVCG}.
For a typical visualization application, any embellishment of $\widetilde{\mathbb{Z}}_1$ would incur more cost for additional processing.
In other words, as shown in Figure \ref{fig:Training}, users should be able to react more quickly or even more accurately if $\widetilde{\mathbb{Z}}_1$ is pre-processed in the direction towards $\widetilde{\mathbb{Z}}_n$.
However, the practical experience suggests that the demand is to embellish $\widetilde{\mathbb{Z}}_1$ with more realism in many dimensions of immersion and presence.

The reason for this apparent contradiction is that VE-based training is a form of \emph{model-developmental visualization} and the actual alphabets $\mathbb{Z}_1, \mathbb{Z}_2, \ldots, \mathbb{Z}_n$ along a visualization process has to include the \emph{model} being developed.
When one is developing a machine-centric model, such as a decision tree in \cite{Tam:2017:TVCG}, the initial alphabet $\mathbb{Z}_1$ encompasses all possible variations of the decision tree model in the context ($\mathbb{M}$), all possible variations of the inputs to the model ($\mathbb{I}$), and all possible variations of the outputs to the model ($\mathbb{O}$).
At the beginning of the workflow, we do not know how these three components are related to each other.
So $\mathbb{Z}_1$ has the highest level of uncertainty as it encompasses all combinations of three types of variations $\mathbb{Z}_1 = \mathbb{M} \times \mathbb{I} \times \mathbb{O}$.
Through a visualization-assisted learning workflow, we gradually narrow down a specific model and establish the functional relationships among the three variations, which is represented by $\mathbb{Z}_n$ at the end of the process.
Typically $\mathbb{Z}_n$ encompasses only one model, or a few models.
For simplicity, let us make $m_{best} \in \mathbb{M}$ as the chosen optimal model.
It can be written as $\mathbb{Z}_n = \{ [m_{best}, i, o] \, | \, o = m_{best}(i), m_{best} \in \mathbb{M}, i \in \mathbb{I}, o \in \mathbb{O} \} )$.
Since the number of letters in $\mathbb{Z}_1$ is at the scale of $||\mathbb{M}|| \cdot ||\mathbb{I}|| \cdot ||\mathbb{O}||$, and that of $\mathbb{Z}_n$ is at the scale of $||\mathbb{I}||$, $\mathbb{Z}_n$ has a much lower entropy than $\mathbb{Z}_1$.
Hence the trend of alphabet compression is consistent with the theory proposed in \cite{Chen:2016:TVCG}.

The mathematical formulation of a VE-based training system is fundamentally the same as that of the aforementioned visualization-assisted learning workflow.
The main difference is that we cannot directly visualize a human-centric model (e.g., the brain function for controlling a type of motion), at least currently, but we can normally do so for a machine-centric centric model (e.g., a decision tree).
Because in a real-world environment, a human-centric model $m_{best} \in \mathbb{M}$ may require more complex and detailed inputs (e.g., what type of building, where the fires, etc.) than some abstract information (e.g., 30\% of a building is on fire), making $\mathbb{Z}_1 = \mathbb{V}$ closer to the reality $\mathbb{R}$ partly reflects our attempt to gain the knowledge as to the inputs that the model may depend on.
Without the definite knowledge about these inputs, the best one can do is to provide as much immersion and presence as possible.

\section{Evidence from Cognitive Science}
\label{sec:Psychology}
In this section, we draw evidence from cognitive science to support the theoretical discussions in Section \ref{sec:Theory}.
In particular, we examine the aspects of \emph{attention}, \emph{visual search}, \emph{working memory}, and \emph{motor coordination}. The most relevant findings in cognitive science are detailed in Appendix \ref{app:Psychology}. 

\noindent\textsf{\textbf{Attention}}.
The evidence in cognitive science shows that attention or selective attention is essential for humans to make efficient and effective use of the limited cognitive resources available to each individual \cite{Anderson:2004:book}.
The fine coordination of eye, hand and body movements provide objective details about the organization of attention, working memory and sensorimotor control \cite{hayhoe2005eye}.

For a large display in a VE, participants have to adjust their gaze as well as move their heads.
When participants are at a relatively closer proximity to the display, walking around also becomes necessary.
These additional movements also incur additional requirements for information retention.
Hence, there is a high cognitive load for maintaining a certain level of awareness across the external information available.
For disseminative visualization, a VE system attracts and demands more attention from participants, and can potentially facilitate the delivery of more information for educational purposes.
For observational and analytical visualization, on the other hand, such a demand has to be carefully managed.
The more cognitive resources are devoted to the attention for retrieving external information, the less cognitive resources are available for the attention to internal events (e.g., analytical reasoning and decision making). 

\noindent\textsf{\textbf{Visual Search} and \textbf{Working Memory}}.
Humans are efficient visual searchers.
Cognitive studies confirmed humans' ability to understand a visual scene at a glance \cite{oliva2006building}.
Retention, on the other hand, is not our strength. Humans' short-term (verbal) memory is famously limited to around seven items \cite{miller1956magical}.

Most visualization techniques provide an effective means for external memorization, and utilize our ability in visual search to compensate for limited working memory resources.
In ``big data'' visualization applications, a high-resolution display can provide more display bandwidth for external memorization and enable visual search tasks with less interactions than a low-resolution display.
On the other hand, any humans' soft knowledge about the ``big data'', including the previous visualization experience of the data, is retained through long-term memory, which does not have the same limitation as working memory.
When such knowledge is utilized for visual search, selective attention becomes more effective.
If the high resolution of a display is achieved by a very large display surface, the demand for more cognitive load related to attention may undermine the benefit of visual search with less interaction.

In real-time mixed reality applications, the challenge of the \emph{match} dimension is often related to visual search and memorization.
The integrated presentation of two types of visual stimuli (i.e., virtual and real objects) is not what one encounters in everyday life.
Hence, this unfamiliarity may reduce humans' aforementioned visual search capability.
There can be mismatch between the integrated visualization and the user's mental models gained from real-life experience.
Any mismatch between the two types of visual stimuli (e.g., due to poor registration) can create further difficulties.
Hence, the solutions to these issue include (i) an improvement of the match between the two types of stimuli in order to reduce the user's cognitive load for ``mental registration'' during visual search, and (ii) introducing training in order to improve the relevant mental models of the user retained in the long-term memory.

\noindent\textsf{\textbf{Motor Coordination}}.
One lesson from the past 50 years or so of literature is that moving our bodies is one of the most demanding tasks we perform as humans.
The number of variables in human movement control is estimated to be about $2^{600}$, with considerably simplified  assumptions about motor activations \cite{wolpert2000computational}.

The evidence in cognitive science confirms that the ``models'' of humans' motor coordination are highly complex.
In order for users to develop the ``lost'' motor coordination skills (e.g., due to medical conditions) or some ``new'' skills (e.g., to perform tasks beyond one's natural ability), there is a need for model-developmental visualization.
The use of VEs with a high level of immersion and presence provides more stimulus information to a variety of the variables of a model under training.
This also provides opportunities to researchers in developing the understanding of such a model and its main variables.

\section{Evidence from Practical Applications}
\label{sec:Applications}
Visualization has been a ubiquitous tool for supporting scientific and scholarly activities in almost all disciplines.
Many visualization applications have been developed to run in VEs.
These include applications in
education and e-learning (e.g., \cite{Bell:2016:book}),
design and testing (e.g., \cite{Ong:2004:book}),
sports training (e.g., \cite{Miles:2012:CG}),
information visualization (e.g., \cite{Reda:2013:CGA,Muller:2016:IJSI}),
medicine and healthcare (e.g., \cite{Alaraj:2011:SNI,Cosentino:2013:JRP,Weiss:2016:book}),
environmental planning (e.g., \cite{Papadopoulos:2015:CGA,Papadopoulos:2013:TVCG}),
information dissemination and public engagement (e.g., \cite{Bock:2015:SciVisP}), and
culture and heritage (e.g., \cite{Addison:2000:M}).
In this section, we examine several visualization applications in VEs, and discuss the cost-benefit of such applications based on the experience reported in the literature.
More detailed descriptions and the cost-benefit analysis of the example case studies can be found in Appendix \ref{app:Applications}.

\noindent\textsf{\textbf{Data Visualization on Large Displays}}.
Many empirical studies were carried out to evaluate the utility of large displays for visualization \cite{Bowman:1997:I3D,Yost:2006:TVCG,Jakobsen:2011:CHI,Ball:2007:CHI,Ruddle:2013:IV,Liu:2014:CHI,Reda:2015:CHI,Muller:2016:IJSI}, resulting in a mixed set of conclusions about the relative merits of such VE systems.
Moorland \cite{Moreland:2012:LDAV} summarized a set of challenges in delivering effective visualization on large displays.

M\"{u}ller et al. \cite{Muller:2016:IJSI} reported an empirical study on using large high-resolution displays for comparative visualization.
It is an unbiased piece of investigation into the effectiveness of using large displays (or powerwalls).
They compared a large display (6m $\times$ 2.2m, 10,800 $\times$ 4,096 pixels) with a 24-inch desktop monitor.
They examined visualization tasks for judging the geometric differences among 40 biological structures.
The results of the study showed that accuracy and response times did not differ significantly between different devices.
Participants did not have clear preference towards the large VE display or the desktop monitor.
In such a case, the desktop monitor was seen as a more economical choice.

From the perspective of information-theoretic cost-benefit analysis, we can observe that the visualization task was to examine the relationship amongst 40 data objects, and is at the level of analytical visualization.
Because the total number of possible relationships is relative low (780), the task was carried out with brute-force observation, in other words, more similar to typical observational visualization.
The task has a well-defined decision alphabet, and hence the alphabet compression is substantial.
The dependent variables (e.g., accuracy and response time) of the study relate directly to the potential distortion and cognitive cost in the cost-benefit metric.
From the perspective of cognitive science, the visualization task is a relatively complex visual search task, and demands working memory to retain some interim comparative judgements.
Hence any additional head and body movement may incur more cognitive load.
In their results, there is a small trend of high response time for the large display, which might indicate such extra load.
Meanwhile, the benefit of the large higher resolution display is unclear as participants viewed two types of displays at different distances.
The requirement for display resolution is also complex for geometrical comparison, as the judgement is likely made at multiple levels of overview and details.

The study indicates that to achieve sufficient cost-benefit of using VEs in observational and analytical visualization for ``big data'' is not trivial.
Nevertheless, once we understood the three abstract measures of alphabet compression, potential distortion, and cost, we can explore this avenue further by considering visualization tasks that may demand more alphabet compression (e.g., relationships among 400 or 4,000 structures), and the need for cost reduction by using some analytics algorithms to prioritize the comparative activities.

\noindent\textsf{\textbf{Surgical Training}}
Domain experts in medicine are early adopters of VEs, particularly in the context of training surgical procedures.
Traditionally surgical training is an apprenticeship model whereby trainees observe the procedure being performed, before attempting it for themselves (under guidance) on real patients.
However, this apprenticeship model is being challenged because of the quality and safety standards in surgical training, reduction in training hours, and constant technological advances.
As a result, pressure on training outside the operating room has significantly increased.
A variety of training aids are available, such as mannequins, but are often unrealistic compared with the real patient.
VE-based training has been widely accepted as a complementary training methodology for well over two decades (e.g., \cite{letterie2002virtual,seymour2008vr,larsen2012efficacy,zendejas2013state}.
Typically a VE helps to develop hand eye coordination and other psychomotor skills, while catering for different patient types and enabling the exploration of what-if scenarios when something goes wrong.

The application of surgical training is a form of model-development visualization.
It places a particular emphasis on vividness and the sense of believing that the virtual patient is real.
The VE alphabet $\mathbb{V}$ encodes the variations of the rendering of the endoscopic view, animation of the virtual patient (e.g., from respiration), and any haptic effect calculated on the virtual endoscope.
The human alphabet $\mathbb{H}$ encodes the variations such as the visual attention of the surgeon, any sensation felt on the surgeon's hands, and the decision on how to proceed from an interpretation of the current state.
The real environment alphabet $\mathbb{R}$ encodes the variations such as the parameter settings on the input interface and the state of the haptic actuator.
The mental models to be trained in such a VE are not only for the surgeon's eye-hand coordination but also for the surgeon's decision mechanism in response to different scenarios.
The cost-benefit of using such VEs has already been confirmed by many practitioners. 

Minimally invasive surgical (MIS) procedures currently provide the most opportunities for surgical training using VEs (see Appendix \ref{sec:Applications} for details).
As MIS can also be deployed in conjunction with real-time mixed reality systems, the visualization tasks involved also fall into the level of observational visualization, as the surgeon needs to observe a variety of data from both the virtual and real environments frequently and at a quick glance, and to make rapid decisions.
It is a research ambition to evolve such systems further to surgical guidance systems to be deployed in real operation rooms.
In other words, there are continuing research effort to increase the space of the real environment alphabet $\mathbb{R}$.
The visualization tasks performed in such surgical guidance systems will be mission-critical, and the necessity for achieving high rate alphabet compression (i.e., from data to decision) with minimal potential distortion will be paramount.

\noindent\textsf{\textbf{Sports Training}}.
Sporting activities can lend themselves very well to being replicated within a VE.
In the context of visualization, domain experts in sports are interested in using VEs to provide alternative ways of training a skill, and analysing performance.
Miles et al. \cite{miles2012review} provide a comprehensive review of the use of VEs for training in ball sports.
They identified several key research challenges, including:
what technologies achieve the best results;
should stereoscopy be used and is a high fidelity VE always better;
what types of skills appear to be best suited to training in VEs; and
whether sports skills reliably transfer from VE training conditions to real-world scenarios?

Many challenges highlighted in \cite{miles2012review} relate to different dimensions of immersion and presence.
For example, the necessity of ``closer approximation of the target skill and the environmental conditions of the target context'' reflects the need to simulate as much reality as possible.
From the perspective of cognitive science, such requirements reflect the complexity of the human models for motor coordination.
The emphasis on ``specific motor control skills'' (e.g., ball passing in rugby \cite{Miles:2014:TCS}) enables the reduction of the complexity of the variable space through domain experts' understand about what may affect such skills.
In other words, this facilitates the reduction of the complexity of the VE alphabet, and thereby the reduction of the cost of using such visualization in a VE.
In addition, the discussions in \cite{miles2012review} on the relative merits of stereoscopic displays and the necessity of high fidelity imagery also reflect the need to understand variable space of individual models under training.
While stereoscopic displays introduce depth perception as a variable in the training of a model, it may also introduce new variables (e.g., fatigue and discomfort, view distortion) that are undesirable to be part of the model.
During a training session, a player processes visual stimuli at a very high speed, achieving extremely high rate of alphabet compression.
Hence the challenge about image fidelity is about how much compression is done by the computer (in the case of low fidelity) and how much is done by humans (in the case of high fidelity).

Miles et al. \cite{Miles:2014:TCS} reported a VE-system for training ball passing skills in rugby as shown in Figure \ref{fig:Examples}(d).
The system simulates a number of variables, such as the flight trajectory of the virtual ball, and wind direction and strength.
They noted that the use of stereoscopy made no significant difference to the accuracy of depth perception in this simulation.
This is a typical visualization task in model development.
Similar to visualization-assisted machine learning \cite{Tam:2017:TVCG}, it is necessary to monitor the variable space of a model, and to relate the performance of the model with various initial conditions.
For VE-based training, the visualization capability is readily available on site.
It is highly desirable to utilize such capability for supporting the model development.

\section{Four Levels of Visualization in VEs}
\label{sec:Vision}
Visualization tasks can be categorized into four levels according to the complexity of their search space \cite{Chen:2016:TVCG}.
In this section, we summarize our theoretical findings at each level, while providing our remarks (indicated by $\blacktriangle$) on new technical challenges.

\noindent\textsf{\textbf{Level 1: Disseminative Visualization}}.
At this level, visualization serves as a presentational aid for disseminating information or insight to others.
While the visualization providers do not purposefully search for new information in the data, it is desirable for the participants at the receiving end to gain as much information as possible.
For a visualization provider, the complexity of the search space is thus $O(1)$, where $O()$ is the big-O notation in complexity analysis.
VEs can be used to maximize the attention of the participants through several dimensions of immersion and presence (e.g., \emph{inclusion}, \emph{surrounding}, \emph{vividness}, and \emph{sense of believing}).
From an information-theoretic perspective, the benefit is achieved primarily through the reduction of potential distortion from the originally intended information, rather than through alphabet compression.
(Otherwise, one would choose to deliver the intended information, for instance, through a list of bullet points.)
Such VEs have a huge value in education and public engagement.
There is a high infrastructural and operational cost to the providers and a high cognitive load to the participants.
There must be continuing provision for the former as many VEs in the categories are providing excellent services to knowledge dissemination.
The latter is incentivized by the novel experience to be gained by the participants, balanced by the demand for attention in an educational process, and rewarded by the amount of information delivered in the process.

$\blacktriangle$ In addition to the financial costs, these VEs continuously face the challenges in delivering technical innovation and novel content. The need for accommodating a large audience is often in conflict with some dimensions of immersion and presence that emphasize the experience of individuals and small groups of participants. 

\noindent\textsf{\textbf{Level 2: Observational Visualization}}.
At this level, visualization enables intuitive and/or speedy observation of captured data.
The complexity of the search space is at the level $O(n)$, where $n$ is the number of data objects.
For visualization tasks involving observing a large amount of data, VEs equipped with large high resolution displays can bring more advantages to applications where datasets are less familiar to the users and there are routine requirements for observing such data (e.g., \cite{Ip:2011:TVCG}).
Such applications demand high-rate alphabet compression and low-rate potential distortion in almost every visualization session.
The better utilization of humans' visual search capability and the provision of higher capacity of external memorization can potentially offset the higher costs than the commodity display screens.

$\blacktriangle$ Our theoretical analysis suggests that it will be helpful to reduce the cognitive load caused by the frequent switching of attention across a wider field of view.
A significant amount of head and body movement for enabling such switching also adds additional addition burden to already-limited working memory.
We therefore hypothesize that medium size high-resolution displays may facilitate the reduction of such cognitive load.
Further studies will be necessary to measure the cognitive loads related to displays of different sizes and different resolutions, and the impact of different levels of fidelity in modelling and rendering.

For visualization tasks to be performed on mixed reality systems, our theoretical analysis confirms the necessity for utilizing parts of reality to reduce the costs and difficulties in capturing, processing, modelling, and rendering many objects in real-time in a real-world environment.
Many mixed reality applications feature datasets that are unfamiliar to the users and requirements for rapid transformation from data to visualization, and then to decision making.
Hence, the conditions for visualization tasks to benefit from VEs are similar to those for the class of ``big data'' applications.

$\blacktriangle$ We recognize that the dimension of \emph{match} poses a major technical challenge.
We have identified the extra cognitive load in visual search due to unfamiliar visual representations and possible poor registration between virtual and real stimuli.
We acknowledge that the existing mixed reality research has already made great effort in improving the accuracy of registration.
We recommend reducing the cognitive load due to unfamiliar representation through innovative design of more ``familiar'' visual representations and introducing necessary training in improving the familiarity.

\noindent\textsf{\textbf{Level 3: Analytical Visualization}}.
At this level, visualization is an investigative aid for examining and understanding complex relationships (e.g., correlation, association, causality, and contradiction).
The complexity of the search space for relationships is typically at the level $O(n^k) (k \geq 2)$, where $n$ is the number of data objects, and $k$ indicates that up to $k$ data objects may be involved in a relation.
Our study of the literature has not revealed any reports of successful deployment of VEs for such visualization tasks.
In general, the more relationships there are to be observed, the more pixels will be required.
However, visualizing a large number of connections across a large display would inevitably introduce a huge amount of cognitive load due to more head and body movement in visual search and more burdens on the limited working memory.

$\blacktriangle$ The lack of concrete evidence does not imply that it is not feasible to use VEs to support analytical visualization.
The high cognitive load in VEs does not imply low cognitive load with commodity computers and displays. 
Once we understand the challenge of the cognitive costs, we may be able to develop new visual representations and visualization techniques that can be effectively deployed  in VEs.
We believe that analytics-aided comparative visualization and visualization-aided causality analysis and predictive analytics are amongst those areas which may yield successful innovation, development, and deployment.

\noindent\textsf{\textbf{Level 4: Model-developmental Visualization}}.
At this level, visualization is a developmental aid for improving existing models, methods, algorithms and systems, as well as for creating new ones.
In this work, we have identified that VE-based training is a form of model-development, though the previous categorization in \cite{Chen:2016:TVCG} considered only machine-centric models.
The complexity of the search space for models is likely to be at the level of NP (non-deterministic, polynomial).
The evidence in cognitive science suggests that human-centric models for motor coordination are more complex than most, if not all, current machine-centric models.
Our theoretical analysis confirms the cost-benefit of VE-based training, and the evidence from practical applications also supports this finding overwhelmingly.

$\blacktriangle$ There are continuing technical challenges to bring more reality into virtuality.
While we develop new techniques to increase the dimensions of immersion and presence, we must also use model-developmental visualization to aid our understanding of the variables in the individual human-centric model under training.
The more understanding we gain, the more effective visualization that we can develop for VE-based training.

$\blacktriangle$ Meanwhile, the use of VEs for developing machine-centric models is yet to be explored.
The successful applications in training human-centric models suggest this potential.
In addition to using VEs to control the visual stimuli for a machine-centric model, we can also potentially observe the evolution of a complex model such as a large neural network in a VE.

\vspace{-2mm}
\section{Answering Practical Questions}
\label{sec:QA}
So far we have shown that the cost-benefit analysis based on information theory can explain why different types of VEs have different impacts on each of the levels of visualization tasks, and such explanations can be supported by evidences from cognitive science and practical applications. If the above theoretical discourse is correct, we should also expect the cost-benefit analysis can be applied to practical problems that have not yet been solved. While there will be a journey from any theory to a corresponding practical solution, the theory should at least offer an effective a pathway to a solution.

As part of IEEE VIS 2017, the attendees of the \emph{Workshop on Immersive Analytics: Exploring Future Interaction and Visualization Technologies for Data Analytics} (\url{(http://immersiveanalytics.net/}). posed a number of questions for discussions during the Workshop. Since the discussions on many questions were largely from a practical perspective and often inconclusive, they offer an opportunity to test the usefulness of the cost-benefit analysis based on information theory. There a total of 16 questions. As detailed in Appendix \ref{app:QA}, we have attempted the answers to 13 of these questions. Here we use our answers to Q1 and Q11 as examples to demonstrate that the cost-benefit analysis can offer an effective pathway to help advance the discourse.

\noindent\textbf{Q1.} \emph{Immersion: How immersive is too immersive?}\\
\indent To formulate an answer to this question, one needs to consider what immersion is and how its quantity is estimated. This paper answers the first question by making use of the existing definitions of the dimensions of VEs (Section 3), and answers the second question by introducing information-theoretic measures to visualization processes in different types of virtual environments (Section 4). The amount of immersion is reflected by the amount of Shannon entropy of the virtual environment and that of the real environment experienced by participants. In addition, we can also measure the amount of Shannon entropy of the data space $\mathbb{Z}_1$ to be visualized and the complexity of visualization tasks $\mathbb{Z}_n$. The paper examines how positive and negative impact of immersion and presence in four categories of VE systems (Section 4) and different levels of visualization (Section 7). One way to consider \textbf{Q1} is to rephrase the question as how to optimize the cost-benefit of immersion. The theoretical answer is summarized in Table 1 and Section 7, while in practice we can use the similar discourse in Sections 4 and 6 to analyze a practical application

\noindent\textbf{Q11.} \emph{Do we really need 3D visualization for 3D data?}\\
\indent We assume that the term ``3D visualization'' implies the use of a 3D volumetric display or a 2D stereo display. This question is indeed at the heart of the cost-benefit analysis. Let us compare the process for generating a visualization alphabet on a 3D visualization environment with the process involving a plain 2D environment. For the same 3D data alphabet, the former is likely to result in less \emph{Alphabet compression}, less \emph{Potential Distortion}, less cognitive \emph{Cost}, but more economic \emph{Cost}. The \emph{Potential Distortion} and cognitive \emph{Cost} in the reverse mapping from the visualization alphabet to the data alphabet depends partly on the viewer's knowledge about the data being visualized. If a viewer is familiar with the variations in the data alphabet, such as different chairs, the \emph{Potential Distortion} and cognitive \emph{Cost} can be very similar between the two types of visualization environments. Hence, the higher \emph{Alphabet Compression} and lower economic \emph{Cost} in the plain 2D environment can bring more cost-benefit. On the other hand, if the variations in the data alphabet is unfamiliar to the viewer, such as the swarming shapes of a large school of fish, the plain 2D environment will likely result in more \emph{Potential Distortion} and cognitive \emph{Cost}. Here we use the term ``alphabet'' throughout the discussion to emphasize that we are not considering only a single dataset rather all possible datasets that a viewer can encounter in a particular context.

Hence, the question does not have a yes or no answer, but an optimization solution based on the cost-benefit metric. In addition, we also need to look forward to the decision alphabet following the visualization process. Some types of potential distortion (e.g., the shape of individual fish) may have less impact on the decision about the collective shape of schooling fish. In such a scenario, one may ask if using a gigapixel display would bring much more benefit than an original desktop display. Similarly, one can also apply the analysis to compare 3D geometric models displayed as outlines, wireframe, shaded, and photorealistic objects using a 2D display.

\vspace{-2mm}
\section{Conclusions}
\label{sec:Conclusions}
In this paper, we have applied information theory in general, and the recently proposed cost-benefit model \cite{Chen:2016:TVCG} in particular, to an array of visualization tasks in VEs.
The cost-benefit analysis allows us to examine different aspects of VEs and visualization in abstraction, and to make generalized observations.
The evidence from cognitive science supports our analysis of various cognitive costs in VEs, and the evidence from practical applications substantiates the benefits of using VEs for visualization in conditions suggested by the theoretical analysis.
We believe that this theoretical study has resulted in several contributions.
It provides an objective assessment of the cost-benefit of visualization in VEs, and presents a set of theory-informed recommendations for future development in this area.
It extends the original definition of four levels of visualization, and validates the cost-benefit metric through its application to a large research area intersecting visualization and VEs.
We hope that many researchers, including ourselves, will explore various challenges presented in Section \ref{sec:Vision}, while seizing the opportunity of continuing reduction of the cost of some VE devices. 



\bibliographystyle{eg-alpha-doi}

\bibliography{Theory-VR}

\newpage

~\\

\newpage

\appendix

\section{\textbf{Further Details on Evidence from Cognitive Science}}
\label{app:Psychology}
In this section, we draw evidence from cognitive science to support the theoretical discussions in Section \ref{sec:Theory}.
In particular, we examine the aspects of \emph{attention}, \emph{visual search}, \emph{working memory}, and \emph{motor coordination}.
We use $\blacktriangleright$ at the beginning of a paragraph to indicate our observations and remarks.

\vspace{1mm}
\noindent\textsf{\textbf{Attention}}.
Attention is a complex cognitive function that selects an aspect of external information (e.g., visual, audio, smell, etc.) or internal events (e.g., thoughts) and maintains a certain level of awareness.
Attention or selective attention is essential for humans to make efficient and effective use of the limited cognitive resources available to each individual \cite{Anderson:2004:book}.
The anatomy of the human eye reflects the compromises necessary in applying limited attentional resources to varying task demands.
The foveated structure of the eye imposes a substantial constraint on the human visual system.
We can only physically direct our gaze, and consequently a huge proportion of the \emph{neural resources} in our visual system \cite{VanEssen1984}, towards one small area of visual space at a time.
To compensate for this, we have evolved sophisticated selective visual attention circuitry allowing us to rapidly redeploy these neural resources as necessary \cite{desimone1995neural}.

Eye movements are the direct consequence of shifts in \emph{overt attention}.
The fine coordination of eye, hand and body movements provide objective details about the organization of attention, working memory and sensorimotor control \cite{hayhoe2005eye}.
Eye movements reflect information retrieval relevant to the current visual task \cite{Yarbus1967}.
The mechanical costs of eye movements are inconsequential \cite{robinson1964mechanics}.
We make many gaze shifts during our day to day activities with varying magnitude, timing, and apparent purpose \cite{land2001ways}.

$\blacktriangleright$ For a large display in a VE, participants have to adjust their gaze as well as move their heads.
When participants are at a relatively closer proximity to the display, walking around also becomes necessary.
These additional movements also incur additional requirements for information retention.
Hence, there is a high cognitive load for maintaining a certain level of awareness across the external information available.
For disseminative visualization, a VE system attracts and demands more attention from participants, and can potentially facilitate the delivery of more information for educational purposes.
For observational and analytical visualization, on the other hand, such a demand has to be carefully managed.
The more cognitive resources are devoted to the attention for retrieving external information, the less cognitive resources are available for the attention to internal events (e.g., analytical reasoning and decision making). 

\vspace{1mm}
\noindent\textsf{\textbf{Visual Search}}.
Humans are efficient visual searchers.
Cognitive studies confirmed humans' ability
to understand a visual scene at a glance \cite{oliva2006building},
to search for known signals embedded in visual noise \cite{najemnik2005optimal},
to identify outlier targets rapidly \cite{treisman1980feature,wolfe1994guided},
to take advantage of spatial cuing \cite{posner1980attention}, and
to predict probable locations for targets \cite{torralba2006contextual}.

\vspace{1mm}
\noindent\textsf{\textbf{Working Memory}}.
Retention, on the other hand, is not our strength.
Humans' short-term (verbal) memory is famously limited to around seven items \cite{miller1956magical}.
Modern theory emphasizes the importance of \emph{working memory} on cognitive tasks \cite{baddeley1974working}, \cite{baddeley1992working}.
Working memory includes both visual and phonological (verbal) components, mirroring perceptual modalities.
The capacity of visual working memory is difficult to measure precisely.
It has been estimated that we can store a conjunction of features representing about four discrete objects \cite{luck1997capacity}.
More recently, information theoretic models implying a flexibly allocated capacity account for behavior better than models with a fixed number of slots \cite{sims2012ideal}.
Regardless, there is general agreement that working memory is a highly constrained resource.

The limits of visual working memory are more apparent in what we miss than in what we retain.
In the phenomenon of \emph{change blindness} \cite{simons1997change}, \cite{rensink2002change} large objects in a scene can be introduced, changed, or completely removed without an observer being aware.
Visual awareness of the change is masked with a short visual interruption such as a flash, cut, or eye movement \cite{o1999change}.
Change blindness is the consequence of selective attention and allocation of limited working memory resources.

$\blacktriangleright$ Most visualization techniques provide an effective means for external memorization, and utilize our ability in visual search to compensate for limited working memory resources.
In ``big data'' visualization applications, a high-resolution display can provide more display bandwidth for external memorization and enable visual search tasks with less interactions than a low-resolution display.
On the other hand, any humans' soft knowledge about the ``big data'', including the previous visualization experience of the data, is retained through long-term memory, which does not have the same limitation as working memory.
When such knowledge is utilized for visual search, selective attention becomes more effective.
If the high resolution of a display is achieved by a very large display surface, the demand for more cognitive load related to attention may undermine the benefit of visual search with less interaction.

$\blacktriangleright$ In real-time mixed reality applications, the challenge of the \emph{match} dimension is often related to visual search and memorization.
The integrated presentation of two types of visual stimuli (i.e., virtual and real objects) is not what one encounters in everyday life.
Hence, this unfamiliarity may reduce humans' aforementioned visual search capability.
There can be mismatch between the integrated visualization and the user's mental models gained from real-life experience.
Any mismatch between the two types of visual stimuli (e.g., due to poor registration) can create further difficulties.
Hence, the solutions to these issue include (i) an improvement of the match between the two types of stimuli in order to reduce the user's cognitive load for ``mental registration'' during visual search, and (ii) introducing training in order to improve the relevant mental models of the user retained in the long-term memory.

\vspace{1mm}
\noindent\textsf{\textbf{Motor Coordination}}.
One lesson from the past 50 years or so of literature is that moving our bodies is one of the most demanding tasks we perform as humans.
We typically control only very specific task relevant dimensions.
Despite considerable variability across a huge number of kinematic degrees of freedom, the error in a blacksmith's strike point is measured in mm \cite{Bernstein67}.
Optimal feedback control seems to provide a compelling mathematical account of this sort of minimization of error along task-relevant dimensions potentially at the expense of increased variability along task-irrelevant dimensions \cite{Todorov2002}.
The number of variables in human movement control is estimated to be about $2^{600}$, with considerably simplified  assumptions about motor activations \cite{wolpert2000computational}.
This is more than the number of atoms in the universe.

Cognitive studies have confirmed many fascinating properties of humans' motor coordination.
These include eye-head-hand coordination with precise timing \cite{pelz2001coordination},
peripheral monitoring of the position of the finger and making corrections to match the intended trajectory \cite{saunders2004visual}, 
taking into consideration our own intrinsic motor variability \cite{trommershauser2003statistical} and environmental variability \cite{seydell2008learning}, and
making look-ahead fixations to improve the accuracy of grasping \cite{mennie2007look,brouwer2007role}.
The spatiotemporal complexity of humans' motor coordination challenges any attempt to define a model accurately.
For example, visual references to stepping locations are at least two steps ahead in order for humans to maintain an efficient walking gait \cite{matthis2014visual}.
The spatiotemporal coordination of eye and body movements is tightly controlled.
Pointing gaze towards a location in space commits a huge proportion of neural resources to that area.
There is a blurry line between high-level motor planning areas and high-level sensory, attention and association areas. There is a continuous remapping of sensory and remembered information into a manual, or at least motor-centric mapping of space \cite{graziano1998spatial,graziano2002cortical}.
Organizing sensorimotor control is one of, if not the most important function of cerebral cortex.
Despite the mathematical complexity, our brains have been optimized by evolution to solve this particular problem very efficiently.

$\blacktriangleright$ The evidence in cognitive science confirms that the ``models'' of humans' motor coordination are highly complex.
In order for users to develop the ``lost'' motor coordination skills (e.g., due to medical conditions) or some ``new'' skills (e.g., to perform tasks beyond one's natural ability), there is a need for model-developmental visualization.
The use of VEs with a high level of immersion and presence provides more stimulus information to a variety of the variables of a model under training.
This also provides opportunities to researchers in developing the understanding of such a model and its main variables.

\section{\textbf{Further Details on Evidence from Practical Applications}}
\label{app:Applications}
Visualization has been a ubiquitous tool for supporting scientific and scholarly activities in almost all disciplines.
Many visualization applications have been developed to run in VEs.
These include applications in
education and e-learning (e.g., \cite{Bell:2016:book}),
design and testing (e.g., \cite{Ong:2004:book}),
sports training (e.g., \cite{Miles:2012:CG}),
information visualization (e.g., \cite{Reda:2013:CGA,Muller:2016:IJSI}),
medicine and healthcare (e.g., \cite{Alaraj:2011:SNI,Cosentino:2013:JRP,Weiss:2016:book}),
environmental planning (e.g., \cite{Papadopoulos:2015:CGA,Papadopoulos:2013:TVCG}),
information dissemination and public engagement (e.g., \cite{Bock:2015:SciVisP}), and
culture and heritage (e.g., \cite{Addison:2000:M}).
In this section, we examine several visualization applications in VEs, and discuss the cost-benefit of such applications based on the experience reported in the literature.
Similarly, we use $\blacktriangleright$ at the beginning of a paragraph to indicate our observations and remarks.

\vspace{1mm}
\noindent\textsf{\textbf{Data Visualization on Large Displays}}.
Many empirical studies were carried out to evaluate the utility of large displays for visualization \cite{Bowman:1997:I3D,Yost:2006:TVCG,Jakobsen:2011:CHI,Ball:2007:CHI,Ruddle:2013:IV,Liu:2014:CHI,Reda:2015:CHI,Muller:2016:IJSI}, resulting in a mixed set of conclusions about the relative merits of such VE systems.
Moorland \cite{Moreland:2012:LDAV} summarized a number of observations about the challenges in delivering effective visualization on large displays.

M\"{u}ller et al. \cite{Muller:2016:IJSI} reported an empirical study on using large high-resolution displays for comparative visualization.
It is an unbiased piece of investigation into the effectiveness of using large displays (or powerwalls).
They compared a large display (6m $\times$ 2.2m, 10,800 $\times$ 4,096 pixels) with a 24-inch desktop monitor.
They examined visualization tasks for judging the geometric differences among 40 biological structures.
The results of the study showed that accuracy and response times did not differ significantly between different devices.
Participants did not have clear preference towards the large VE display or the desktop monitor.
In such a case, the desktop monitor was seen as a more economical choice.

$\blacktriangleright$ From the perspective of information-theoretic cost-benefit analysis, we can observe that the visualization task was to examine the relationship amongst 40 data objects, and is at the level of analytical visualization.
Because the total number of possible relationships is relative low (780), the task was carried out with brute-force observation, in other words, more similar to typical observational visualization.
The task has a well-defined decision alphabet, and hence the alphabet compression is substantial.
The dependent variables (e.g., accuracy and response time) of the study relate directly to the potential distortion and cognitive cost in the cost-benefit metric.
From the perspective of cognitive science, the visualization task is a relatively complex visual search task, and demands working memory to retain some interim comparative judgements.
Hence any additional head and body movement may incur more cognitive load.
In their results, there is a small trend of high response time for the large display, which might indicate such extra load.
Meanwhile, the benefit of the large higher resolution display is unclear as participants viewed two types of displays at different distances.
The requirement for display resolution is also complex for geometrical comparison, as the judgement is likely made at multiple levels of overview and details.

$\blacktriangleright$ The study indicates that to achieve sufficient cost-benefit of using VEs in observational and analytical visualization for ``big data'' is not trivial.
Nevertheless, once we understood the three abstract measures of alphabet compression, potential distortion, and cost, we can explore this avenue further by considering visualization tasks that may demand more alphabet compression (e.g., relationships among 400 or 4,000 structures), and the need for cost reduction by using some analytics algorithms to prioritize the comparative activities.

\vspace{1mm}
\noindent\textsf{\textbf{Surgical Training}}
Domain experts in medicine are early adopters of VEs, particularly in the context of training surgical procedures.
Traditionally surgical training is an apprenticeship model whereby trainees observe the procedure being performed, before attempting it for themselves (under guidance) on real patients.
However, this apprenticeship model is being challenged because of the quality and safety standards in surgical training, reduction in training hours, and constant technological advances.
As a result, pressure on training outside the operating room has significantly increased.
A variety of training aids are available, such as mannequins, but are often unrealistic compared with the real patient.
VE-based training has been widely accepted as a complementary training methodology for well over two decades (e.g., \cite{letterie2002virtual,seymour2008vr,larsen2012efficacy,zendejas2013state}.
Typically a VE helps to develop hand eye coordination and other psychomotor skills, while catering for different patient types and enabling the exploration of what-if scenarios when something goes wrong.

Minimally invasive surgical (MIS) procedures currently provide the most opportunities for surgical training using VEs, and several commercial systems are available from companies such as 3D Systems Healthcare (CO, USA) and Mentice (Gothenburg, Sweden).
MIS procedures may be within the abdominal or pelvic cavities (laparoscopy) or the thoracic or chest cavity (thoracoscopy).
They are typically performed far from the target location through small incisions elsewhere in the body.
The surgeon's view of the patient is limited to the endoscopic camera view displayed on a monitor.
Mixed reality MIS systems are currently being developed for operating theatres, whereby the endoscope camera view is augmented with other information that may not be visible. 
Haptic feedback on the laparoscopic tools, e.g., the endoscope, may provide the surgeon with additional cues.
A processing flow for a typical MIS trainer using the forwarding connections defined in Figure \ref{fig:ProcessFlow} is: 

\textcircled{\small{1}} Endoscope virtual camera position has changed; virtual endoscopic view on computer monitor is updated and re-rendered.

\textcircled{\small{2}} Endoscope virtual camera position has changed; Surgeon interprets current view and decides on next step (e.g., insertion, retraction, perform biopsy).

\textcircled{\small{4}} Surgeon decides to manipulate endoscope interface moving the endoscope within the virtual patient; Surgeon interprets new view from the endoscopic camera (perhaps in conjunction with medical scan images).

\textcircled{\small{5}} Surgeon decides to manipulate endoscope interface, to move the endoscope within the virtual patient; virtual endoscopic view on computer monitor is updated and re-rendered.

\textcircled{\small{7}} A setting changes on the input interface hardware (which is typically fabricated to look and feel like a real endoscopic device); Output interfaces (computer monitor, but could be a head mounted display (HMD); actuator inside input interface hardware provides tactile or force cue) are updated.

$\blacktriangleright$ The application of surgical training is a form of model-development visualization.
It places a particular emphasis on vividness and the sense of believing that the virtual patient is real.
The VE alphabet $\mathbb{V}$ encodes the variations of the rendering of the endoscopic view, animation of the virtual patient (e.g., from respiration), and any haptic effect calculated on the virtual endoscope.
The human alphabet $\mathbb{H}$ encodes the variations such as the visual attention of the surgeon, any sensation felt on the surgeon's hands, and the decision on how to proceed from an interpretation of the current state.
The real environment alphabet $\mathbb{R}$ encodes the variations such as the parameter settings on the input interface and the state of the haptic actuator.
The mental models to be trained in such a VE are not only for the surgeon's eye-hand coordination but also for the surgeon's decision mechanism in response to different scenarios.
The cost-benefit of using such VEs has already been confirmed by many practitioners. 

$\blacktriangleright$ As MIS can also be deployed in conjunction with real-time mixed reality systems, the visualization tasks involved also fall into the level of observational visualization, as the surgeon needs to observe a variety of data from both the virtual and real environments frequently and at a quick glance, and to make rapid decisions.
It is a research ambition to evolve such systems further to surgical guidance systems to be deployed in real operation rooms.
In other words, there are continuing research effort to increase the space of the real environment alphabet $\mathbb{R}$.
The visualization tasks performed in such surgical guidance systems will be mission-critical, and the necessity for achieving high rate alphabet compression (i.e., from data to decision) with minimal potential distortion will be paramount.

\vspace{1mm}
\noindent\textsf{\textbf{Sports Training}}.
Sporting activities can lend themselves very well to being replicated within a VE.
This could be purely for entertainment purposes such as golf and basketball simulators found in arcades, or non-immersive computer games on popular games consoles.
In the context of visualization, domain experts in sports are interested in using VEs to provide alternative ways of training a skill, and analysing performance.
Miles et al. \cite{miles2012review} provide a comprehensive review of the use of VEs for training in ball sports.
They identify the key research challenges that are currently being explored, including:
what technologies achieve the best results;
should stereoscopy be used and is a high fidelity VE always better;
what types of skills appear to be best suited to training in VEs; and
whether sports skills reliably transfer from VE training conditions to real-world scenarios?
The broad coverage of this review and its objective assessment the current successes and challenges can provide our theoretical analysis with necessary evidence in practical applications.

Closely related to the topic of this review, Miles et al. \cite{Miles:2014:TCS} reported a VE-system for training ball passing skills in rugby as shown in Figure \ref{fig:Examples}(d).
The system simulates a number of variables, such as the flight trajectory of the virtual ball, and wind direction and strength.
They also noted that the use of stereoscopy made no significant difference to the accuracy of depth perception in this simulation.

$\blacktriangleright$ Many challenges highlighted in \cite{miles2012review} relate to different dimensions of immersion and presence.
For example, the necessity of ``closer approximation of the target skill and the environmental conditions of the target context'' reflects the need to simulate as much reality as possible.
From the perspective of cognitive science, such requirements reflect the complexity of the human models for motor coordination.
The emphasis on ``specific motor control skills'' (e.g., ball passing in rugby \cite{Miles:2014:TCS}) enables the reduction of the complexity of the variable space through domain experts' understand about what may affect such skills.
In other words, this facilitates the reduction of the complexity of the VE alphabet, and thereby the reduction of the cost of using such visualization in a VE.
In addition, the discussions in \cite{miles2012review} on the relative merits of stereoscopic displays and the necessity of high fidelity imagery also reflect the need to understand variable space of individual models under training.
While stereoscopic displays introduce depth perception as a variable in the training of a model, it may also introduce new variables (e.g., fatigue and discomfort, view distortion) that are undesirable to be part of the model.
During a training session, a player processes visual stimuli at a very high speed, achieving extremely high rate of alphabet compression.
Hence the challenge about image fidelity is about how much compression is done by the computer (in the case of low fidelity) and how much is done by humans (in the case of high fidelity).

$\blacktriangleright$ Miles et al. \cite{miles2012review} pointed out the importance of performance measure and analysis in VE-based training.
This is a typical visualization task in model development.
Similar to visualization-assisted machine learning \cite{Tam:2017:TVCG}, it is necessary to monitor the variable space of a model, and to relate the performance of the model with various initial conditions.
For VE-based training, the visualization capability is readily available on site.
It is highly desirable to utilize such capability for supporting the model development.

\section{\textbf{Can the Theory Answer Practical Questions?}}
\label{app:QA}

As part of IEEE VIS 2017, the attendees of the \emph{Workshop on Immersive Analytics: Exploring Future Interaction and Visualization Technologies for Data Analytics} (\url{(http://immersiveanalytics.net/}). posed a number of questions for discussions during the Workshop. As the discussions on many questions were largely from a practical perspective and often inconclusive, here we attempt the answers to thirteen of these questions primarily using the information-theoretic metric for measuring the cost-benefit of visualization in VEs. Although the theory cannot fully answer all questions, as demonstrated below, it can help advance the discourse significantly.

Note that the Q8 was missing in the original table in the Google document \url{ https://goo.gl/d5pbRG}. We omitted Q12 and Q15 (about designing empirical studies) and Q13 (about existing design methodologies), because they are beyond the scope of this paper. To accommodate different lengths of questions and answers, we reformatted the 16 questions slightly by changing from a table to a list. We also removed the names of those who proposed the questions.

\subsection{Q1. Immersion: How immersive is too immersive?}
To formulate an answer to this question, one needs to consider what immersion is and how its quantity is estimated. This paper answers the first question by making use of the existing definitions of the dimensions of VEs (Section 3), and answers the second question by introducing information-theoretic measures to visualization processes in different types of virtual environments (Section 4). The amount of immersion is reflected by the amount of Shannon entropy of the virtual environment and that of the real environment experienced by participants. In addition, we can also measure the amount of Shannon entropy of the data space $\mathbb{Z}_1$ to be visualized and the complexity of visualization tasks $\mathbb{Z}_n$. The paper examines how positive and negative impact of immersion and presence in four categories of VE systems (Section 4) and different levels of visualization (Section 7). One way to consider this question is to rephrase the question as how to optimize the cost-benefit of immersion. The theoretical answer is summarized in Table 1 and Section 7, while in practice we can use the similar discourse in Sections 4 and 6 to analyze a practical application.

\subsection{Q2. Walking: During immersive visual exploration, do we walk or do we sit? Do we walk around the data or through the data?}
These two questions can only be answered properly after considering the specific type of data, their possible explicit or metaphoric representations in VEs, and the likely availability of the users' \emph{a priori} knowledge about the data. The dimension \emph{match} is particularly important to the first question. The second question relates to the theoretic discussion about the visual information-seeking mantra in \cite{Chen:2010:TVCG,Chen:2016:book}. Information-theoretically, if the user does not have a holistic mental model about the data and such a model is useful for performing the visualization tasks, an initial well-designed walk-around can have a similar effect as an overview, first which is shown to be cost-beneficial \cite{Chen:2010:TVCG}. If the user already has a good mental model about the data or such a mental model does not benefit the visualization tasks to be performed, a walk-through the data is likely to have more cost-benefit \cite{Chen:2016:book}.

\subsection{Q3. Abstract Data: Why do we need immersive visualization for non-spatial data? How can we immerse into non-spatial data?}
This question relates to the discussions in Sections 4.1 and 4.3, the first case study in Section 6, and the discussions on analytical visualization and model-developmental visualization for machine-centric models in Section 7.

\subsection{Q4. Experiential Analytics: How do we understand and design for this experience? When is it essential and for whom? Are there counter examples where it is unnecessary and slows down the analytical process?}
The first two questions correspond to the discussion on analytical visualization in Section 7. The empirical study by M\"{u}ller et al. \cite{Muller:2016:IJSI} discussed in Section 6 relates to the third question. Clearly much more research effort will be required to answer these three questions.

\subsection{Q5. Engagement and Attraction: Immersiveness for engagement (only)? What makes us feel immersed, what do we connect to?}
We believe that this paper has provided detailed answers to the first two questions. Here we assume that the third question means ``the connection between a VE and our mind''. Section 5 and Appendix A provide a summary answer to this question.

\subsection{Q6. Immersive vs 3D: How does immersive analytics differ from 3D data visualization? Non-3D immersive visualization? Most of papers show virtual environments (data visualization) but no data analytics. How we can actually analyze data within immersive environments as we can do in a 2D desktop interface?}
Spatially-3D data visualization can be carried out using immersive and semi-immersive VEs as well as using non-immersive display environments. For the questions about data analytics, see \textbf{Q3} and \textbf{Q4}.

\subsection{Q7. Mapping 3D geospatial datasets into real-world VR environments: how does the quality of the environment impact the understanding of results?}
As discussed in this paper, the impact depends partly on the visualization task, and we can start to examine the impact by first determining which level of visualization the task resides at. For example, the discussions in Sections 4.1 and 7 are particularly relevant to disseminative visualization, while the discussions in Sections 4.2, 4.3, and 7 are relevant to observational visualization.

\subsection{Q9. Interaction: Which interaction modalities would you pick? What about mixing modalities of interaction? What are sensible combinations? What can be used to build passive/proactive context and how can that context be used in more explicit/reactive interactions? Which visualization tasks are applicable to immersive analytics?}
This paper provides limited coverage on the interaction modalities. This is partly because the application of information theory to interaction in conventional visualization environments is not yet well addressed. We hope that future research effort into building a theoretical foundation of visualization will bring compressive answers to these questions.

\subsection{Q10. What does it mean to create a visualization in Immersive Analytics?}
In our information-theoretic model of visualization, a visualization process is a series transformation of visualization alphabets, which is part of the other three alphabets (see the paragraphs under the heading of Alphabets and Letters in Section 4). For example, a digital dataset may be represented by a graphical representation in a VE alphabet and a contextual ``dataset'' may be a part of a reality alphabet in a mixed-reality environment. The events observed and the decisions made by a user are likely to be part of the human alphabet.

\subsection{Q11. Do we really need 3D visualization for 3D data? (related to Q3) What can we perceive/do in 3D immersion that cannot be perceived/done with 2D representations? (related to Q6)}
Here we assume that the term ``3D visualization'' implies the use of a 3D volumetric display device or a 2D stereo display device. This question is indeed at the heart of the cost-benefit analysis. Let us compare the process for generating a visualization alphabet on a 3D visualization environment with the process involving a plain 2D environment. For the same 3D data alphabet, the former is likely to result in less \emph{Alphabet compression}, less \emph{Potential Distortion}, less cognitive \emph{Cost}, but more economic \emph{Cost}. The \emph{Potential Distortion} and cognitive \emph{Cost} in the reverse mapping from the visualization alphabet to the data alphabet depends partly on the viewer's knowledge about the data being visualized. If a viewer is familiar with the variations in the data alphabet, such as different chairs, the \emph{Potential Distortion} and cognitive \emph{Cost} can be very similar between the two types of visualization environments. Hence, the higher \emph{Alphabet Compression} and lower economic \emph{Cost} in the plain 2D environment can bring more cost-benefit. On the other hand, if the variations in the data alphabet is unfamiliar to the viewer, such as the swarming shapes of a large school of fish, the plain 2D environment will likely result in more \emph{Potential Distortion} and cognitive \emph{Cost}. Here we use the term ``alphabet'' throughout the discussion to emphasize that we are not considering only a single dataset rather all possible datasets that a viewer can encounter in a particular context.

Hence, the question does not have a yes or no answer, but an optimization solution based on the cost-benefit metric. In addition, we also need to look forward to the decision alphabet following the visualization process. Some types of potential distortion (e.g., the shape of individual fish) may have less impact on the decision about the collective shape of schooling fish. In such a scenario, one may ask if using a gigapixel display would bring much more benefit than an original desktop display. Similarly, one can also apply the same analysis to compare 3D geometric models displayed as outlines, wireframe, shaded, and photorealistic objects using a plain 2D display.

See also the answers to \textbf{Q3}, \textbf{Q4}, and \textbf{Q6}.

\subsection{Q14. Are ``classical'' definitions (Milgram and Kishino's, and Azuma's) of MR and AR too graphics-centric for data vis? Should we look into more ``experience'' flavors of MR/AR interpretations?}
We think that the questioner is rightly to suggest the need to accommodate ``experience'' in formulating concepts in VEs. Because it is difficult to measure experience and knowledge, the cost-benefit metric proposed in \cite{Chen:2016:TVCG} avoided direct modeling of experience and knowledge. Instead, a user's observations and decisions are explicitly in the alphabets in a data intelligence workflow, while a user's knowledge is implicitly modeled in the reverse mapping function. We believe that more research effort will be necessary for studying the questions in \textbf{Q14}.

\subsection{Q16. Does immersion in data differ from immersion in 3D models? If so should we change how we measure it?}
Normally 3D models, such as volumetric objects in volume rendering and mesh models in surface rendering, are also considered to be datasets. We suspect that the questioner used the term ``data'' to imply datasets with fewer than three spatial dimensions. As the questioner must have already observed, for the datasets with fewer than three spatial dimensions, one would often map some non-spatial variables to unused spatial dimensions (e.g., population to height or time to depth). Such a visual mapping is not uncommon in 2D visualization. For example, the $y$-dimension of a bar chart is commonly used to depict a non-spatial variable. With the aid of other visual variations, more than one non-spatial variable can use the $y$-dimension, e.g., an error bar on top of a height bar. Regardless of whether using spatial or non-spatial models, 2D or 3D visual representations, and VEs or conventional displays, the user has to perform the reverse mapping from a visual channel to a data variable. This reverse mapping always requires some cognitive load and may cause potential distortion. Hence, the information-theoretic metric for the cost-benefit analysis accommodates both forward and backward mappings in visualization processes, and is ideal for comparing the two types of datasets in VEs. On the one hand, based on the notion of \emph{match} discussed in this paper, some well-designed visual mappings from non-spatial data to spatial dimensions may have a good \emph{match} and demand little cognitive load. On the other hand, some real-world 3D models can be unfamiliar to users, and these datasets can still incur undesired potential distortion and cognitive load. So the question cannot be trivially answered based on spatial or non-spatial data.

\subsection{Q17. Defining immersion/immersive. I've heard these hints at a definition:
(1) Immersion has to do with the experience. The person using a system is immersed in the process of analysing data. This, I think, relates to being in flow, and blocking out outside disturbances. Is there a difference between feeling immersed and being immersed? This might be thought about as immersed in analysis.
(2) Immersion has to do with the technology, putting a focus on AR/VR.This might be thought about as the body being immersed.
(3) Immersion has to with being inside/between the data as opposed to looking at it from the outside. This might be thought about as immersed in data.
(4) Immersion has to do with being the social context.}
We hope that the questioner may find the definitions about the dimensions of VEs (Section 3) useful basis for improving the definitions proposed in \textbf{Q17}.

\end{document}